\documentclass[floatfix,nofootinbib,showpacs,showkeys,preprintnumbers,eqsecnum]{revtex4}
\usepackage{dcolumn}%
\usepackage{bm}%
\usepackage{fullpage}
\usepackage[centertags]{amsmath}
\allowdisplaybreaks[1]
\usepackage{amsbsy}
\usepackage{amsfonts}
\usepackage{amssymb}
\usepackage{eufrak}
\newcommand{\vet}[1]{\ensuremath{\hskip-1pt\vec{\hskip1pt#1}}}
\begin{document}
\preprint{hep-ph/0205014}
\title{
Neutrino Wave Packets in Quantum Field Theory
}
\author{C. Giunti}
\email{giunti@to.infn.it}
\homepage{http://www.to.infn.it/~giunti}
\affiliation{
INFN, Sezione di Torino,
and
Dipartimento di Fisica Teorica, Universit\`a di Torino,\\
Via P. Giuria 1, I--10125 Torino, Italy
}
\date{26 June 2002}
\begin{abstract}
We present a model of neutrino oscillations
in the framework of quantum field theory
in which the
propagating neutrino and
the particles participating to the production and detection
processes are described by wave packets.
The neutrino state
is a superposition
of massive neutrino wave packets determined by the production
process, as naturally expected from causality.
We show that the energies and momenta of the
massive neutrino components
relevant for neutrino oscillations
are in general different
from the average energies and momenta of the
propagating massive neutrino wave packets,
because of the effects of the detection process.
Our results confirm the correctness of
the standard expression for the oscillation length
of extremely relativistic neutrinos
and the existence of a coherence length.
\end{abstract}
\maketitle

\section{Introduction}
\label{Introduction}

The theory of neutrino oscillations
is an important field of research,
especially
after the convincing evidence of the existence of neutrino oscillations
obtained in atmospheric and solar neutrino experiments
(see \cite{Kajita:2000mr,Ahmad:2001an,Fogli:2001vr,Giunti:2001ws,Bahcall:2001pe,Ahmad:2002jz,Ahmad:2002ka}).

Neutrinos can oscillate if neutrino mixing is realized in nature
(see \cite{Bilenky-Pontecorvo-PR-78,%
Bilenky-Petcov-RMP-87,%
CWKim-book-93,%
BGG-review-98}),
\emph{i.e.}
if the left-handed components
$\nu_{{\alpha}L}$
of the flavor neutrino fields
($\alpha=e,\mu,\tau$)
are superpositions of the left-handed components
$\nu_{aL}$
of the fields of neutrinos with definite mass:
\begin{equation}
\nu_{{\alpha}L}
=
\sum_a
U_{{\alpha}a}
\,
\nu_{aL}
\,,
\label{mixing}
\end{equation}
where $U$ is the neutrino mixing matrix.
Here the field
$\nu_a$
describes a neutrino with mass $m_a$.
A flavor neutrino produced in a charged-current weak interaction process
is a superposition of
massive neutrino components with well defined kinematical properties.
Neutrino oscillations
are space-dependent and/or time-dependent flavor transitions generated by
the different phase velocities of different massive neutrino components.
The probabilities of
flavor transitions in neutrino oscillation experiments
depend on the differences
$\Delta{m}^2_{ab} \equiv m_a^2 - m_b^2$
of the squared neutrino masses
and on the elements of the mixing matrix $U$.

Different models of neutrino oscillations
have been presented by several authors
(see \cite{CWKim-book-93,Zralek:1998rp,Beuthe:2001rc,Neutrino_Unbound}
and references therein).
In this paper we adopt a wave packet approach,
in the line of Refs.~\cite{Giunti-Kim-Lee-Whendo-91,Giunti-Kim-Coherence-98,%
Giunti-Kim-Lee-Lee-93,Giunti-Kim-Lee-Whendo-98}.

As discussed in Ref.~\cite{Kayser-oscillations-81},
neutrino oscillations are possible
only
if the processes of neutrino production and detection
have momentum uncertainties
that allow the coherent production and detection of different
massive neutrino components.
From the fundamental principles of quantum mechanics it follows that
the production and detection processes must be localized
and
the massive neutrino components of a flavor neutrino
must be described by wave packets.
We have developed a quantum-mechanical
wave-packet description of neutrino oscillations
in Refs.~\cite{Giunti-Kim-Lee-Whendo-91,Giunti-Kim-Coherence-98}.
This model provides important insights in the
physics of neutrino oscillations.
The main one is the confirmation of the correctness of the standard
expression for the oscillation phase of extremely relativistic neutrinos
(see \cite{Bilenky-Pontecorvo-PR-78,Giunti:2000kw,Bilenky:2001yh,Giunti:2002ee}).
In addition, the quantum-mechanical model of neutrino oscillations
implies the existence of a coherence length,
which was predicted in Ref.~\cite{Nussinov-coherence-76}
on the basis of intuitive quantum-mechanical arguments.
For source--detector distances larger than the coherence length
the flavor transition probability
does not oscillate,
because the separation of the different massive neutrino wave packets,
which propagate with different group velocities,
is so large that they cannot be absorbed coherently in the detection process
\cite{Kiers-Nussinov-Weiss-PRD53-96}.
In this case the probability of flavor-changing transitions
is constant and depends only on the elements of the mixing matrix $U$.

In the quantum-mechanical model of neutrino oscillations
the expression of the state describing a flavor neutrino
as a superposition of massive neutrino components
has to be assumed,
because quantum mechanics is not sufficient for the description
of the neutrino production and detection processes.
Therefore,
it is necessary to develop a description of neutrino oscillations
in the framework of quantum field theory,
which allows to calculate the amplitudes
of neutrino production and detection.
Moreover,
one must notice that the theoretical prediction of
neutrino oscillations
follows from the mixing of neutrino fields in Eq.~(\ref{mixing})
and a consistent theory of neutrino mixing and oscillations
can be formulated only in the framework of quantum field theory.

In Refs.~\cite{Giunti-Kim-Lee-Lee-93,Giunti-Kim-Lee-Whendo-98}
we have developed
a quantum-field-theoretical model of neutrino oscillations
in which the particles taking part to the
neutrino production and detection processes
are described by localized wave packets
and the neutrino propagating between
the production and detection processes is a virtual particle
(see also Refs.~\cite{Cardall-Coherence-99,Beuthe:2002ej} for similar approaches,
Ref.~\cite{Grimus-Stockinger-96,Grimus:1998uh,Grimus:1999ra}
for another quantum-field-theoretical model of neutrino oscillations
with virtual neutrinos,
and Refs.~\cite{Blasone-Vitiello-99,Fujii:2001zv}
for a different quantum-field-theoretical point of view).
This model confirms the correctness of
the standard
expression for the oscillation phase of extremely relativistic neutrinos
and the existence of a coherence length.

However,
since
in oscillation experiments
neutrinos propagate over macroscopic or even astronomical distances,
we think that it is unnatural to consider them as virtual particles,
with undefined properties
(see Refs.~\cite{Giunti-Kim-Lee-Whendo-98,Beuthe:2001rc}).
Since massive neutrinos propagate as free particles
between production and detection,
it should be possible to describe their superposition
constituting the flavor neutrino created in the production process
by an appropriate quantum-field-theoretical state,
similar to the quantum-mechanical wave-packet state
in Ref.~\cite{Giunti-Kim-Coherence-98}.
Causality demands that the neutrino state
is determined by the production process.

In this paper we present a quantum-field-theoretical model
of neutrino oscillations
in which the neutrino propagating between the production and detection
processes is described by a wave packet state
determined by the production process.
This state is derived in
Section~\ref{Production},
starting from a production process
in which the other interacting particles are described by wave packets.
In Section~\ref{Detection}
we calculate the amplitude of the detection process
occurring at a space-time distance
$(\vet{L},T)$
from the production process,
using the neutrino state obtained in Section~\ref{Production} and
wave packets for the other interacting particles.
In Section~\ref{Probability}
we calculate the probability of neutrino oscillations in space
from the average over the unmeasured propagation time $T$
of the detection probability
and we discuss the effects of the detection
process on neutrino oscillations.
Conclusions are presented in Section~\ref{Conclusions}.

In the following we assume
that the particles that participate to the production and detection
processes can be described by appropriate wave packets.
This is possible if their properties are determined.
If the information about their properties
is incomplete,
as often occurs in practice,
each particle must be described by a statistical operator,
also known as ``density matrix'',
constructed from an incoherent mixture of wave packets
with definite properties.
As a consequence,
also the propagating neutrino must be described by a statistical operator
constructed from an incoherent mixture of the pure states that we derive
in
Section~\ref{Production},
and the oscillation probability is given by an appropriate average
of the oscillation probability that we derive in
Section~\ref{Probability}
over the unknown properties of the particles
participating to the production and detection
processes.

\section{Production}
\label{Production}

The approach presented here is based on the
fact that in quantum field theory
the effect of interactions is described
by the operator $\mathcal{S}-\mathbf{1}$,
with
\begin{equation}
\mathcal{S}
=
T
\,
\exp\left(
- i \int \mathrm{d}^4x \,
\mathcal{H}_I(x)
\right)
\,,
\label{S}
\end{equation}
where 
$\mathcal{H}_I(x)$
is the interaction Hamiltonian
expressed in terms of the appropriate field operators.
Given
an asymptotic initial state
$ | i \rangle $
the asymptotic final state
resulting from an interaction is given by
\begin{equation}
| f \rangle
\propto
(\mathcal{S}-\mathbf{1})
| i \rangle
\simeq
- i \int \mathrm{d}^4x \,
\mathcal{H}_I(x) \,
| i \rangle
\,,
\label{fti}
\end{equation}
where the last expression applies
to first order in perturbation theory,
which we adopt in the following,
because we consider weak interaction processes.
Hence,
in quantum field theory it is possible
to calculate with Eq.(\ref{fti})
the final state resulting from any interaction
and in particular the final state
of the processes in which a neutrino is produced.
Since in the majority of experiments
neutrinos are produced in charged-current weak decays,
we consider the production process
\begin{equation}
P_I \to P_F + \ell^{+}_{\alpha} + \nu_{\alpha}
\,,
\label{prod}
\end{equation}
in which $P_I$ is the decaying particle,
$P_F$ is a decay product (absent in two-body decays)
and $\ell^{+}_{\alpha}$ is the final state charged lepton
that determines the flavor of the produced neutrino $\nu_{\alpha}$.
As a simple example of a possible production process
of the type (\ref{prod})
we will consider the pion decay
\begin{equation}
\pi^{+} \to \mu^{+} + \nu_{\mu}
\,,
\label{pion-decay}
\end{equation}
where
$ P_I = \pi^{+} $,
$P_F$ is absent
and
$ \ell^{+}_{\alpha} = \mu^{+} $.

Let us consider the production process (\ref{prod}).
The final state obtained with Eq.(\ref{fti})
describes all the final particles
of the process in an entangled way:
\begin{equation}
| \widetilde{P}_F , \widetilde{\ell}^{+}_{\alpha} , \widetilde{\nu}_{\alpha} \rangle
\propto
- i \int \mathrm{d}^4x \,
\mathcal{H}_I^P(x) \,
| P_I \rangle
\,,
\label{entangled}
\end{equation}
where
$ | P_I \rangle $
is the state describing the initial particle $P_I$
and
$\mathcal{H}_I^P(x)$
is the Hamiltonian describing the production process.

The entangled state
$ | \widetilde{P}_F , \widetilde{\ell}^{+}_{\alpha} , \widetilde{\nu}_{\alpha} \rangle $
can be disentangled by measuring the properties
of the particles involved.
In particular,
in the study of neutrino oscillations,
one is interested in the knowledge of the state
$ | \nu_{\alpha} \rangle $
describing the neutrino produced
in a process of type (\ref{prod}).
In order to disentangle the state
$ | \nu_{\alpha} \rangle $,
it is necessary to measure the properties of the
other particle in the final state,
\emph{i.e.} of $P_F$ and $\ell^{+}_{\alpha}$.
This measurement does not have to be done necessarily
by a specific instrument,
but could be done by the interactions
of the anti-lepton
$\ell^{+}_{\alpha}$
and of
$P_F$
with the surrounding medium.
The measurement process
causes a collapse of the entangled final state
$ | \widetilde{P}_F , \widetilde{\ell}^{+}_{\alpha} , \widetilde{\nu}_{\alpha} \rangle $
to a disentangled state
$
| P_F \rangle
| \ell^{+}_{\alpha} \rangle
| \nu_{\alpha} \rangle
$,
with
the neutrino state
$ | \nu_{\alpha} \rangle $
given by
\begin{equation}
| \nu_{\alpha} \rangle
\propto
\left(
\langle P_F |
\langle \ell^{+}_{\alpha} |
\right)
| \widetilde{P}_F , \widetilde{\ell}^{+}_{\alpha} , \widetilde{\nu}_{\alpha} \rangle
\propto
\left(
\langle P_F |
\langle \ell^{+}_{\alpha} |
\right)
- i \int \mathrm{d}^4x \,
\mathcal{H}_I^P(x) \,
| P_I \rangle
\,.
\label{state}
\end{equation}

The effective interaction Hamiltonian
that describes the process (\ref{prod})
in the Standard Model is
\begin{eqnarray}
\mathcal{H}_I^P(x)
& = &
\frac{G_F}{\sqrt{2}}
\,
\bar\nu_{\alpha}(x)
\,
\gamma^{\rho}
\left( 1 - \gamma_5 \right)
\ell_{\alpha}(x)
\,
J^{P}_{\rho}(x)
\nonumber
\\
& = &
\frac{G_F}{\sqrt{2}}
\sum_a
U^{*}_{{\alpha}a}
\,
\bar\nu_a(x)
\,
\gamma^{\rho}
\left( 1 - \gamma_5 \right)
\ell_{\alpha}(x)
\,
J^{P}_{\rho}(x)
\,,
\label{HIP}
\end{eqnarray}
where
$G_F$ is the Fermi constant
and
$J^{P}_{\rho}(x)$
is the weak charged current
that describes the transition
$ P_{I} \to P_{F} $.

In order to simplify as much as possible our discussion,
let us consider the process (\ref{prod})
with the particles
$ P_I $, $ P_F $ and $ \ell^{+}_{\alpha} $
described by the wave-packet states
\begin{equation}
| \chi \rangle
=
\int \mathrm{d}^3p
\,
\psi_{\chi}(\vet{p};\vet{p}_{\chi},\sigma_{p\chi})
| \chi(\vet{p},h_{\chi}) \rangle
\,,
\label{states}
\end{equation}
where
$\chi = P_I , P_F , \ell^{+}_{\alpha}$,
the momentum distributions are denoted by
$\psi_{\chi}(\vet{p};\vet{p}_{\chi},\sigma_{p\chi})$,
and
$h_{\chi}$ are the helicities.
We assume that the particles
$ P_I $, $ P_F $ and $ \ell^{+}_{\alpha} $
are polarized.
If the measurement process is not sufficient to determine the polarization
of these particles,
each of them must be described by a statistical operator
(density matrix)
constructed from an incoherent mixture of
the pure states (\ref{states}) with different helicities.
Consequently,
the propagating neutrino must be described by a statistical operator
constructed from an incoherent mixture of
the pure states (\ref{state})
obtained with different helicities of the particles
$ P_I $, $ P_F $ and $ \ell^{+}_{\alpha} $.

We consider the Gaussian momentum distributions\footnote{
A Gaussian momentum distribution is the most convenient one
for the calculation of several integrations in the following.
Other distributions which are sharply peaked
around an average momentum $\vet{p}_{\chi}$
lead to the same results after their approximation
with a Gaussian
in order to perform the integrals with a saddle-point approximation.
Therefore,
the Gaussian momentum distributions
can be taken as approximations of the real momentum distributions
from the beginning.
}
\begin{equation}
\psi_{\chi}(\vet{p};\vet{p}_{\chi},\sigma_{p\chi})
=
\left( 2 \pi \sigma_{p\chi}^2 \right)^{-3/4}
\exp\left[
-
\frac
{ ( \vet{p} - \vet{p}_{\chi} )^2 }
{ 4 \sigma_{p\chi}^2 }
\right]
\,,
\label{wf}
\end{equation}
where
$\vet{p}_{\chi}$
and
$\sigma_{p\chi}$
are,
respectively,
the average momentum
and the momentum uncertainty of the wave packet of the particle $\chi$.
The corresponding
wave functions in coordinate space are given by
\begin{equation}
\psi_{\chi}(\vet{x},t;\vet{p}_{\chi},\sigma_{p\chi})
=
\langle 0 | \chi(x) | \chi \rangle
\approx
\int \frac{ \mathrm{d}^3p }{ (2\pi)^{3/2} }
\,
\psi_{\chi}(\vet{p};\vet{p}_{\chi},\sigma_{p\chi})
\,
e^{ - i E_{\chi}(\vet{p}) t + i \vet{p} \vet{x} }
\,,
\label{wfc1}
\end{equation}
where
\begin{equation}
E_{\chi}(\vet{p}) = \sqrt{ \vet{p}^2 + m_{\chi}^2 }
\label{052}
\end{equation}
is the energy corresponding to the momentum $\vet{p}$,
and we have neglected for simplicity
the spin degrees of freedom.
The Gaussian momentum distributions (\ref{wf})
are assumed to be sharply peaked around their average momentum,
i.e.
the condition
$
\sigma_{p\chi}
\ll
E_{\chi}^2(\vet{p}_{\chi}) / m_{\chi}
$
is assumed to be satisfied.
Under this condition,
the energy
$ E_{\chi}(\vet{p}) $
can be approximated by\footnote{
With this approximation we neglect the spreading of the wave packets.
}
\begin{equation}
E_{\chi}(\vet{p})
\simeq
E_{\chi}
+
\vet{v}_{\chi} ( \vet{p} - \vet{p}_{\chi} )
\,,
\label{energy}
\end{equation}
where
\begin{equation}
E_{\chi}
\equiv
E_{\chi}(\vet{p}_{\chi})
=
\sqrt{ \vet{p}_{\chi}^2 + m_{\chi}^2 }
\label{eave}
\end{equation}
is the average energy
(up to corrections of the order $\sigma_{p\chi}^2/E_{\chi}$, see Eq.~(\ref{166}))
and
\begin{equation}
\vet{v}_{\chi}
\equiv
\left.
\frac{ {\partial}E_{\chi} }{ {\partial}\vet{p} }
\right|_{\vet{p}=\vet{p}_{\chi}}
=
\frac
{\vet{p}_{\chi}}
{E_{\chi}}
\end{equation}
is the group velocity
of the wave packet of the particle $\chi$.
Under this approximation
the integration in Eq.~(\ref{wfc1}) is Gaussian and leads to
\begin{equation}
\psi_{\chi}(\vet{x},t;\vet{p}_{\chi},\sigma_{p\chi})
\simeq
\left( 2 \pi \sigma_{x\chi}^2 \right)^{-3/4}
\exp\left[
-
i E_{\chi} t
+
i \vet{p}_{\chi} \cdot \vet{x}
-
\frac
{ \left( \vet{x} - \vet{v}_{\chi} t \right)^2 }
{ 4 \sigma_{x\chi}^2 }
\right]
\,,
\label{wfc2}
\end{equation}
where
$\sigma_{x\chi}$
defined by the relation
\begin{equation}
\sigma_{x\chi} \, \sigma_{p\chi} = \frac{1}{2}
\label{sxchi}
\end{equation}
is the spatial width of the wave packet.
One can see that
the wave functions
overlap at the origin of space-time coordinates,
where the neutrino production process takes place.
Hence,
wave packet states (\ref{states})
are appropriate for the description of the particles
taking part to the localized production process (\ref{prod}).

Let us determine
the energy uncertainty of the wave packet state (\ref{states})
describing a localized particle $\chi$.
Developing $E_{\chi}(\vet{p})$
up to second order in the power series of
$( \vet{p} - \vet{p}_{\chi} )$,
one obtains the average energy\footnote{
I would like to thank M. Beuthe for
pointing out the necessity to develop
$E_{\chi}(\vet{p})$
up to second order in the power series of
$( \vet{p} - \vet{p}_{\chi} )$.
As a consequence,
Eqs.~(\ref{166}), (\ref{167}), (\ref{455}), (\ref{456}) and (\ref{464})
have been corrected with respect to the first version
of the paper appeared in the electronic archive hep-ph,
where
$E_{\chi}(\vet{p})$
was developed only to first order
(as in Eq.~(\ref{energy})).
}
\begin{equation}
\langle E \rangle_{\chi}
=
\langle \chi | \widehat{E} | \chi \rangle
\simeq
E_{\chi}
+
\left( 3 - \vet{v}_{\chi}^2 \right) \frac{\sigma_{p\chi}^2}{2E_{\chi}}
\,,
\label{166}
\end{equation}
where
$\widehat{E}$
is the energy operator.
The average squared energy is given exactly by
\begin{equation}
\langle E^2 \rangle_{\chi}
=
\langle \chi | \widehat{\vet{P}}^2 | \chi \rangle + m_\chi^2
=
E_{\chi}^2
+
3 \, \sigma_{p\chi}^2
\,,
\label{E2chi}
\end{equation}
where
$\widehat{\vet{P}}$
is the momentum operator,
leading to the squared energy uncertainty
\begin{equation}
\langle (\delta{E})^2 \rangle_{\chi}
=
\langle \chi | ( \widehat{E} - E_{\chi} )^2 | \chi \rangle
=
\vet{v}_{\chi}^2 \, \sigma_{p\chi}^2
\,.
\label{167}
\end{equation}
Therefore,
somewhat surprisingly,
a localized particle at rest has momentum uncertainty
without energy uncertainty
(at order $\sigma_{p\chi}/E_{\chi}$).

We use the following
Fourier expansion for the spin $1/2$ fermion fields:
\begin{equation}
\mathfrak{f}(x)
=
\int
\frac
{ \mathrm{d}^3p }
{ (2\pi)^{3/2} }
\sum_{h=\pm1}
\left[
a_{\mathfrak{f}}(\vet{p},h)
\,
u_{\mathfrak{f}}(\vet{p},h)
\,
e^{ - i p x }
+
b_{\mathfrak{f}}^{\dagger}(\vet{p},h)
\,
v_{\mathfrak{f}}(\vet{p},h)
\,
e^{ i p x }
\right]_{p^0=E_{\mathfrak{f}}(\vet{p})}
\,,
\label{fer}
\end{equation}
where
$h$ is the helicity,
$u_{\mathfrak{f}}(\vet{p},h)$
and
$v_{\mathfrak{f}}(\vet{p},h)$
are four-component spinors,
and
$a_{\mathfrak{f}}(\vet{p},h)$ and $b_{\mathfrak{f}}(\vet{p},h)$
are the particle and anti-particle destruction operators
obeying the canonical anticommutation relations
\begin{equation}
\{ a_{\mathfrak{f}}(\vet{p},h) , a_{\mathfrak{f}}^{\dagger}(\vet{p}',h') \}
=
\{ b_{\mathfrak{f}}(\vet{p},h) , b_{\mathfrak{f}}^{\dagger}(\vet{p}',h') \}
=
\delta^3(\vet{p}-\vet{p}') \, \delta_{hh'}
\,.
\label{anticommutation}
\end{equation}
The one-particle fermion states with definite momentum and helicity
$ | \mathfrak{f}(\vet{p},h) \rangle = a_{\mathfrak{f}}^{\dagger}(\vet{p},h) \, | 0 \rangle $
are normalized by the relation
$
\langle \mathfrak{f}(\vet{p},h) | \mathfrak{f}(\vet{p}',h') \rangle
=
\delta^3(\vet{p}-\vet{p}') \, \delta_{hh'}
$.
One can easily check that in this way
the wave-packet states (\ref{states}) for $\chi=\mathfrak{f}$
are normalized to
$ \langle \mathfrak{f} | \mathfrak{f} \rangle = 1 $.

From Eqs.(\ref{state}), (\ref{HIP}) and (\ref{states}),
the state that describes the neutrino produced in the process
(\ref{prod})
is given by
\begin{eqnarray}
| \nu_{\alpha} \rangle
&\propto&
\sum_a U_{{\alpha}a}^{*}
\int
\mathrm{d}^3p'_{P_F}
\,
\psi_{P_F}^{*}(\vet{p}'_{P_F};\vet{p}_{P_F},\sigma_{pP_F})
\int
\mathrm{d}^3p'_{\ell^{+}_{\alpha}}
\,
\psi_{\ell^{+}_{\alpha}}^{*}(\vet{p}'_{\ell^{+}_{\alpha}};\vet{p}_{\ell^{+}_{\alpha}},\sigma_{\ell^{+}_{\alpha}})
\nonumber
\\
&& \null
\times
\int
\mathrm{d}^3p'_{P_I}
\,
\psi_{P_I}(\vet{p}'_{P_I};\vet{p}_{P_I},\sigma_{pP_I})
\nonumber
\\
&& \null
\times
\int
\mathrm{d}^4x
\left(
\langle P_F(\vet{p}'_{P_F}) |
\langle \ell^{+}_{\alpha}(\vet{p}'_{\ell^{+}_{\alpha}}) |
\right)
\bar\nu_a(x)
\,
\gamma^{\rho}
\left( 1 - \gamma_5 \right)
\ell_{\alpha}(x)
\,
J^{P}_{\rho}(x)
\,
| P_I(\vet{p}'_{P_I}) \rangle
\,.
\label{p01}
\end{eqnarray}
Using the Fourier expansion (\ref{fer})
for the lepton fields,
we obtain
\begin{eqnarray}
&& \null
| \nu_{\alpha} \rangle
\propto
\sum_a U_{{\alpha}a}^{*}
\int
\mathrm{d}^4x
\int
\mathrm{d}^3p'_{P_F}
\,
\psi_{P_F}^{*}(\vet{p}'_{P_F};\vet{p}_{P_F},\sigma_{pP_F})
\nonumber
\\
&& \null
\times
\int
\mathrm{d}^3p'_{\ell^{+}_{\alpha}}
\,
\psi_{\ell^{+}_{\alpha}}^{*}(\vet{p}'_{\ell^{+}_{\alpha}};\vet{p}_{\ell^{+}_{\alpha}},\sigma_{\ell^{+}_{\alpha}})
\int
\mathrm{d}^3p'_{P_I}
\,
\psi_{P_I}(\vet{p}'_{P_I};\vet{p}_{P_I},\sigma_{pP_I})
\,
J^{P}_{\rho}(\vet{p}'_{P_F},h_{P_F};\vet{p}'_{P_I},h_{P_I})
\nonumber
\\
&& \null
\times
\int
\mathrm{d}^3p
\sum_{h}
\bar{u}_{\nu_a}(\vet{p},h)
\gamma^{\rho}
\left( 1 - \gamma_5 \right)
v_{\ell^{+}_{\alpha}}(\vet{p}'_{\ell^{+}_{\alpha}},h_{\ell^{+}_{\alpha}})
\,
e^{ i ( p + p'_{\ell^{+}_{\alpha}} + p'_{P_F} - p'_{P_I} ) x }
\,
| \nu_a(\vet{p},h) \rangle
\,,
\label{p02}
\end{eqnarray}
with
$
p^0
=
E_{\nu_a}(\vet{p})
$,
$
p^{\prime0}_{P_I}
=
E_{P_I}(\vet{p}'_{P_I})
$,
$
p^{\prime0}_{P_F}
=
E_{P_F}(\vet{p}'_{P_F})
$,
$
p^{\prime0}_{\ell^{+}_{\alpha}}
=
E_{\ell^{+}_{\alpha}}(\vet{p}'_{\ell^{+}_{\alpha}})
$,
and
$
J^{P}_{\rho}(\vet{p}'_{P_F},h_{P_F};\vet{p}'_{P_I},h_{P_I})
=
\langle P_F(\vet{p}'_{P_F},h_{P_F}) |
\,
J^{P}_{\rho}(0)
\,
| P_I(\vet{p}'_{P_I},h_{P_I}) \rangle
$.
For example,
in the pion decay process (\ref{pion-decay})
$J^{P}_{\rho}(\vet{p}'_{P_F},h_{P_F};\vet{p}'_{P_I},h_{P_I})$
is given by
$
\langle 0 |
\,
J^{P}_{\rho}(0)
\,
| \pi^{+}(\vet{p}'_{\pi^{+}}) \rangle
=
f_{\pi}
\,
(p'_{\pi^{+}})_{\rho}
$,
where
$f_{\pi}$
is the pion decay constant.

Since the wave packets of $P_I$, $P_F$ and $\ell^{+}_{\alpha}$
are assumed to be sharply peaked around the
respective average momenta,
the integrations over
$\mathrm{d}^3p'_{P_F}$,
$\mathrm{d}^3p'_{\ell^{+}_{\alpha}}$
and
$\mathrm{d}^3p'_{P_I}$
can be performed with a saddle-point approximation
using the approximation (\ref{energy}),
which leads to
\begin{eqnarray}
| \nu_{\alpha} \rangle
&\propto&
\sum_a U_{{\alpha}a}^{*}
\int \mathrm{d}^3p
\sum_h
\mathcal{A}^P_a(\vet{p},h)
\,
| \nu_a(\vet{p},h) \rangle
\nonumber
\\
&& \null
\times
\int
\mathrm{d}^4x
\,
\exp\left[
- i ( E_P - E_{\nu_a}(\vet{p}) ) t
+ i ( \vet{p}_P - \vet{p} ) \vet{x}
-
\frac
{ \vet{x}^2
- 2 \, \vet{v}_P \cdot \vet{x} \, t
+ \Sigma_P \, t^2 }
{ 4 \sigma_{xP}^2 }
\right]
\,,
\label{p04}
\end{eqnarray}
with
\begin{eqnarray}
&& \null
E_P
\equiv
E_{P_I}
-
E_{P_F}
-
E_{\ell^{+}_{\alpha}}
\,,
\label{EP}
\\
&& \null
\vet{p}_P 
\equiv
\vet{p}_{P_I}
-
\vet{p}_{P_F}
-
\vet{p}_{\ell^{+}_{\alpha}}
\,,
\label{pP}
\\
&& \null
\frac{1}{\sigma_{xP}^2}
\equiv
\frac{1}{\sigma_{xP_{I}}^2}
+
\frac{1}{\sigma_{xP_{F}}^2}
+
\frac{1}{\sigma_{x\ell^{+}_{\alpha}}^2}
\,,
\label{sxP}
\\
&& \null
\vet{v}_P
\equiv
\sigma_{xP}^2
\left(
\frac{ \vet{v}_{P_I} }{ \sigma_{xP_{I}}^2 }
+
\frac{ \vet{v}_{P_F} }{ \sigma_{xP_{F}}^2 }
+
\frac{ \vet{v}_{\ell^{+}_{\alpha}} }{ \sigma_{x\ell^{+}_{\alpha}}^2 }
\right)
\,,
\label{vP}
\\
&& \null
\Sigma_P
\equiv
\sigma_{xP}^2
\left(
\frac{ \vet{v}_{P_I}^2 }{ \sigma_{xP_{I}}^2 }
+
\frac{ \vet{v}_{P_F}^2 }{ \sigma_{xP_{F}}^2 }
+
\frac{ \vet{v}_{\ell^{+}_{\alpha}}^2 }{ \sigma_{x\ell^{+}_{\alpha}}^2 }
\right)
\,,
\label{SP}
\\
&& \null
\mathcal{A}^P_a(\vet{p},h)
\equiv
\bar{u}_{\nu_a}(\vet{p},h)
\,
\gamma^{\rho}
\left( 1 - \gamma_5 \right)
v_{\ell^{+}_{\alpha}}(\vet{p}_{\ell^{+}_{\alpha}},h_{\ell^{+}_{\alpha}})
\,
J^{P}_{\rho}(\vet{p}'_{P_F},h_{P_F};\vet{p}'_{P_I},h_{P_I})
\,.
\label{APa}
\end{eqnarray}
As naturally expected,
the overall spatial width $\sigma_{xP}$
of the production process is dominated by the smallest
among the spatial widths
of $P_I$, $P_F$ and $\ell^{+}_{\alpha}$.
The quantities $|\vet{v}_P|$ and $\Sigma_P$
are limited by
\begin{equation}
0
\leq
|\vet{v}_P|
\leq
1
\,,
\qquad
0
\leq
\Sigma_P
\leq
1
\,.
\label{431}
\end{equation}

Carrying out the Gaussian integral over $\mathrm{d}^4x$,
we obtain the neutrino state
\begin{equation}
| \nu_{\alpha} \rangle
=
N_{\alpha}
\sum_a U_{{\alpha}a}^{*}
\int
\mathrm{d}^3p
\,
e^{-S^P_a(\vet{p})}
\sum_h
\mathcal{A}^P_a(\vet{p},h)
\,
| \nu_a(\vet{p},h) \rangle
\,,
\label{099}
\end{equation}
where $N_{\alpha}$
is a normalization constant such that
\begin{equation}
\langle \nu_{\alpha} | \nu_{\alpha} \rangle
=
1
\,,
\label{norma}
\end{equation}
and
\begin{equation}
S^P_a(\vet{p})
\equiv
\frac
{ \left( \vet{p}_P - \vet{p} \right)^2 }
{ 4 \, \sigma_{pP}^2 }
+
\frac
{ \left[ E_P - E_{\nu_a}(\vet{p})
-
\left( \vet{p}_P - \vet{p} \right)
\cdot \vet{v}_P \right]^2 }
{ 4 \, \sigma_{pP}^2 \lambda_P }
\,.
\label{SPa}
\end{equation}
Here
\begin{equation}
\lambda_P
\equiv
\Sigma_P - \vet{v}_P^2
\,,
\label{432}
\end{equation}
such that
\begin{equation}
0
\leq
\lambda_P
\leq
1
\,,
\label{433}
\end{equation}
and we have defined the momentum uncertainty $\sigma_{pP}$
through the relation
\begin{equation}
\sigma_{xP} \, \sigma_{pP}
=
\frac{ 1 }{ 2 }
\,.
\label{067}
\end{equation}
The overall momentum uncertainty
is dominated by the largest momentum
uncertainty among $P_I$, $P_F$ and $\ell^{+}_{\alpha}$:
\begin{equation}
\sigma_{pP}^2
=
\sigma_{pP_{I}}^2
+
\sigma_{pP_{F}}^2
+
\sigma_{p\ell^{+}_{\alpha}}^2
\,.
\label{spP}
\end{equation}

The neutrino state (\ref{099})
describes a neutrino produced in the weak interaction process
(\ref{prod})
as a superposition of massive neutrino components.
In Section~\ref{Detection}
we discuss the detection of this state
and in Section~\ref{Probability}
we derive the corresponding transition probability. 
In the following part of this Section
we discuss some properties of the
massive neutrino wave-packet states
\begin{equation}
| \nu_a \rangle
=
N_a
\int
\mathrm{d}^3p
\,
e^{-S^P_a(\vet{p})}
\sum_h
\mathcal{A}^P_a(\vet{p},h)
\,
| \nu_a(\vet{p},h) \rangle
\,,
\label{0991}
\end{equation}
with the normalization constant $N_a$ such that
\begin{equation}
\langle \nu_a | \nu_a \rangle = 1
\,.
\label{450}
\end{equation}
The massive neutrino wave-packet states (\ref{0991})
are the components of the state (\ref{099}),
which can be written as
\begin{equation}
| \nu_{\alpha} \rangle
=
N_{\alpha}
\sum_a 
\frac{U_{{\alpha}a}^{*}}{N_a}
\,
| \nu_a \rangle
\,,
\label{0992}
\end{equation}
and the normalization constant
$N_{\alpha}$
is related to the normalization constants
$N_a$
by
\begin{equation}
N_{\alpha}
=
\left(
\sum_a 
\frac{|U_{{\alpha}a}|^2}{N_a^2}
\right)^{-1/2}
\,.
\label{0993}
\end{equation}

The normalization condition (\ref{450})
requires that
\begin{equation}
N_a^2
\sum_h
\int
\mathrm{d}^3p
\,
|\mathcal{A}^P_a(\vet{p},h)|^2
\,
e^{-2S^P_a(\vet{p})}
=
1
\,.
\label{451}
\end{equation}
The integration over $\mathrm{d}^3p$ can be done with a saddle point approximation
around the stationary point of $S^P_a(\vet{p})$,
\begin{equation}
\left.
\frac{ {\partial}S^P_a(\vet{p}) }{ {\partial}\vet{p} }
\right|_{\vet{p}=\vet{p}_a}
=
0
\,.
\label{peak}
\end{equation}
The momentum $\vet{p}_a$ is given by
\begin{equation}
\vet{p}_P
-
\vet{p}_a
+
\frac{1}{\lambda_P}
\left[
E_P - E_a
-
\left( \vet{p}_P - \vet{p}_a \right) \cdot \vet{v}_P
\right]
\left( \vet{v}_a - \vet{v}_P \right)
=
0
\,,
\label{pnua}
\end{equation}
with
\begin{eqnarray}
&& \null
E_a
\equiv
E_{\nu_a}(\vet{p}_a)
=
\sqrt{ \vet{p}_a^2 + m_a^2 }
\,,
\label{Enua}
\\
&& \null
\vet{v}_a
\equiv
\left.
\frac{ {\partial}E_{\nu_a}(\vet{p}) }{ {\partial}\vet{p} }
\right|_{\vet{p}=\vet{p}_a}
=
\frac{\vet{p}_a}{E_a}
\,.
\label{vnua}
\end{eqnarray}
The saddle-point approximation of the integration over $\mathrm{d}^3p$
in Eq.~(\ref{451})
leads to
\begin{equation}
N_a
=
\left( \frac{ \mathrm{Det}\Lambda_a }{ \pi^3 } \right)^{1/4}
\frac
{ e^{S^P_a(\vet{p}_a)} }
{ \sqrt{ \sum_h |\mathcal{A}^P_a(\vet{p}_a,h)|^2 } }
\,,
\label{452}
\end{equation}
with
\begin{eqnarray}
\Lambda_a^{jk}
&\equiv&
\left.
\frac{ \partial^2 S^P_a(\vet{p}) }{ \partial p^j \partial p^k }
\right|_{\vet{p}=\vet{p}_a}
\nonumber
\\
&=&
\frac{\delta^{jk}}{2\sigma_{pP}^2}
+
\frac{\left(v_P^j-v_a^j\right)\left(v_P^k-v_a^k\right)}{2\sigma_{pP}^2\lambda_P}
-
\frac
{
\left[
\left( E_P - E_a  \right)
-
\left( \vet{p}_P - \vet{p}_a \right) \cdot \vet{v}_P
\right]
}
{2\sigma_{pP}^2\lambda_P}
\,
\frac{\delta^{jk} - v_a^j v_a^k}{E_a}
\,.
\label{123}
\end{eqnarray}

Using the same saddle-point approximation for the integration over $\mathrm{d}^3p$,
one can find that
$\vet{p}_a$
is the average momentum of the state $|\nu_a\rangle$,
\begin{equation}
\langle\vet{p}\rangle_a
=
\langle\nu_a|\widehat{\vet{P}}|\nu_a\rangle
=
\vet{p}_a
\,,
\label{453}
\end{equation}
and the uncertainties of the three momentum components are given by
\begin{equation}
\langle (\delta{p^k})^2 \rangle_a
=
\langle\nu_a| ( \widehat{P}^k - p_a^k )^2 |\nu_a\rangle
=
\frac{1}{2}
\,
(\Lambda_a^{-1})^{kk}
\,.
\label{454}
\end{equation}
Developing $E_{\nu_a}(\vet{p})$
up to second order in the power series of
$( \vet{p} - \vet{p}_a )$
we obtain the average energy
\begin{equation}
\langle E \rangle_a
=
\langle\nu_a|\widehat{E}|\nu_a\rangle
\simeq
E_a
+
\frac
{ \mathrm{Tr}(\Lambda_a^{-1}) - \sum_{j,k} v_a^j (\Lambda_a^{-1})^{jk} v_a^k }
{4E_a}
\,,
\label{455}
\end{equation}
and the energy uncertainty
\begin{equation}
\langle (\delta{E})^2 \rangle_a
=
\langle\nu_a| ( \widehat{E} - E_a )^2 |\nu_a\rangle
\simeq
\frac{1}{2}
\,
\sum_{j,k} v_a^j (\Lambda_a^{-1})^{jk} v_a^k
\,.
\label{456}
\end{equation}

In order to
get further insight in the properties of the massive neutrino
wave-packet states (\ref{0991}),
it is necessary
to solve Eq.~(\ref{pnua})
and determine the values of
$\vet{p}_a$ and $E_a$.
It is important to notice that
the production of the state $|\nu_\alpha\rangle$ in Eq.~(\ref{099})
is not suppressed only if
\begin{equation}
S^P_a(\vet{p}_a)
\lesssim
1
\label{411}
\end{equation}
for all values of the index $a$.
Together with Eq.~(\ref{pnua}),
this inequality constraints the possible values of
$\vet{p}_a$, $\vet{p}_P$ and $E_P$.

Equation~(\ref{pnua})
implies that the massive neutrino momenta
$\vet{p}_a$ must be aligned in the direction
of $\vet{p}_P$:
\begin{equation}
\vet{p}_a = p_a \, \vet{\ell}
\,,
\label{414}
\end{equation}
with
$\vet{p}_P = p_P \, \vet{\ell}$
and
$|\vet{\ell}|=1$.
Hence,
$\vet{\ell}$
is the unit vector
in the direction of propagation of the neutrino.

We solve Eq.~(\ref{pnua}) in the approximation of extremely
relativistic neutrinos.
This approximation is valid in practice because
only neutrinos with energy larger than some fraction of MeV
are detectable.
Indeed,
neutrinos are detected in:

\renewcommand{\labelenumi}{\theenumi.}
\renewcommand{\theenumi}{\arabic{enumi}}

\begin{enumerate}

\item
Charged-current or neutral-current weak processes
which have an energy threshold
larger than some fraction of MeV.
This is due to the fact that
in a scattering process
\begin{equation}
\nu + A \to \sum_X X
\label{500}
\end{equation}
with $A$ at rest,
the squared center-of-mass energy
$s = 2 E_\nu m_A + m_A^2$
(neglecting the neutrino mass)
must be bigger than
$( \sum_X m_X )^2$,
leading to
\begin{equation}
E_{\nu}^{\mathrm{th}}
=
\frac{ ( \sum_X m_X )^2 }{ 2 m_A } - \frac{ m_A }{ 2 }
\,.
\label{501}
\end{equation}
For example:

\begin{itemize}

\item
$
E_{\nu}^{\mathrm{th}}
\simeq
0.233 \, \mathrm{MeV}
$
for
$ \nu_e + {}^{71}\mathrm{Ga} \to {}^{71}\mathrm{Ge} + e^- $
in the GALLEX \cite{GALLEX-99},
SAGE \cite{SAGE-99}
and
GNO \cite{GNO-00}
solar neutrino experiments.

\item
$
E_{\nu}^{\mathrm{th}}
\simeq
0.81 \, \mathrm{MeV}
$
for
$ \nu_e + {}^{37}\mathrm{Cl} \to {}^{37}\mathrm{Ar} + e^- $
in the Homestake \cite{Homestake-98}
solar neutrino experiment.

\item
$
E_{\nu}^{\mathrm{th}}
\simeq
1.8 \, \mathrm{MeV}
$
for
$ \bar\nu_e + p \to n + e^+ $
in reactor neutrino experiments
(for example Bugey \cite{Bugey} and CHOOZ \cite{CHOOZ-99}).

\item
$
E_{\nu}^{\mathrm{th}}
\simeq
2.2 \, \mathrm{MeV}
$
in the neutral-current process
$ \nu + d \to p + n + \nu $
used in the SNO experiment to detect active solar neutrinos
\cite{Ahmad:2002jz}.

\item
$
E_{\nu}^{\mathrm{th}}
\simeq
110 \, \mathrm{MeV}
$
for
$ \nu_\mu + n \to p + \mu^- $.

\item
$
E_{\nu}^{\mathrm{th}}
\simeq
m_\mu^2 / 2 m_e
\simeq
10.9 \, \mathrm{GeV}
$
for
$ \nu_\mu + e^- \to \nu_e + \mu^- $.

\end{itemize}

\item
The elastic scattering process
$ \nu + e^- \to \nu + e^- $,
whose cross section is proportional to the neutrino energy
($
\sigma(E_{\nu})
\sim
\sigma_0 E_{\nu} / m_e
$,
with
$
\sigma_0
\sim
10^{-44} \, \mathrm{cm}^2
$).
Therefore,
an energy threshold
of some MeV's
is needed in order to have a signal above the background.
For example,
$
E_{\nu}^{\mathrm{th}}
\simeq
5 \, \mathrm{MeV}
$
in the Super-Kamiokande \cite{SK-sun-01}
solar neutrino experiment.

\end{enumerate}

Although the direct experimental upper limits
for the effective neutrino masses
in lepton decays are not very stringent
($m_{\nu_e} \lesssim 3 \, \mathrm{eV}$,
$m_{\nu_\mu} \lesssim 190 \, \mathrm{keV}$,
$m_{\nu_\tau} \lesssim 18.2 \, \mathrm{MeV}$,
see Ref.~\cite{PDG}),
we know that
the sum of the masses of the neutrinos
that have a substantial mixing with $\nu_e$, $\nu_\mu$ and $\nu_\tau$
is constrained to be smaller than a few eV
by their contribution to the total energy density of the Universe
\cite{Wang:2001gy,Elgaroy:2002bi,astro-ph/0205223}.

The comparison
of the cosmological limit on neutrino masses
with the energy threshold in the processes of neutrino detection
implies that
the main massive neutrino components of
detectable flavor neutrinos are extremely relativistic\footnote{
It is still possible that
the three active flavor neutrinos
$\nu_e$,
$\nu_\mu$,
$\nu_\tau$
have very small mixing with heavy massive neutrinos that are not relativistic
in some experiment.
In this case,
the heavy neutrino masses
must be taken into account in the calculation of the
production
and
detection
rates of the heavy massive neutrinos,
but the oscillations generated by the large mass differences
between light and heavy neutrinos
are too fast to be observable.
Therefore,
in practice it is sufficient to consider an incoherent mixture of light and heavy
massive neutrinos
that generate a constant flavor-changing transition probability.
Also the mass differences between possible heavy neutrinos
are expected to be too large to generate observable oscillations.
Therefore,
in the following we study the oscillations
due to the mixing
of the flavor neutrinos
$\nu_e$, $\nu_\mu$ and $\nu_\tau$
with
light extremely relativistic massive neutrinos.
}.

We write the average massive neutrino energies $E_a$
in the relativistic approximation\footnote{ \label{degenerate}
In principle
it is possible to consider the approximation of almost degenerate
but not relativistic neutrinos \cite{Beuthe:2002ej}.
This approximation could be relevant in practice
only if
the three active flavor neutrinos
$\nu_e$, $\nu_\mu$ and $\nu_\tau$
mix with heavy and almost degenerate
massive neutrinos,
such that the small mass differences among heavy neutrinos
generate observable oscillations.
Since this possibility seems very unlikely,
we do not consider the case of
almost degenerate
but not relativistic massive neutrinos.
Notice, however,
that a similar approximation is important in the
analogous quantum-field-theoretical treatment of meson oscillations
(see \cite{Beuthe:2001rc}).
}
as
\begin{equation}
E_a \simeq E + \xi \, \frac{m_a^2}{2E}
\,,
\label{101}
\end{equation}
where $E$ is the neutrino energy in the limit of zero mass.
The corresponding momentum has modulus
\begin{equation}
p_a \simeq E - \left( 1 - \xi \right) \frac{m_a^2}{2E}
\,.
\label{102}
\end{equation}

In the following we consider
\begin{equation}
\vet{p}_P = E_P \, \vet{\ell}
\,,
\label{415}
\end{equation}
which
corresponds to exact energy and momentum conservation
in the production process
in the case of massless neutrinos.
It is possible to consider deviations
from Eq.~(\ref{415}) compatible with the inequality
(\ref{411}),
but such deviations would entail
a considerable complication of the formalism
without further insights in the physical properties
of neutrinos emitted in the process (\ref{prod}).

At zeroth order in $m_a^2/E^2$,
Eq.~(\ref{pnua}) imply
\begin{equation}
E = E_P
\,.
\label{106}
\end{equation}
The solution of Eq.~(\ref{pnua})
at first order in $m_a^2/E^2$
gives the value of $\xi$:
\begin{equation}
\xi
=
\frac
{
\lambda_P
-
\vet{\ell} \cdot \vet{v}_P
\left( 1 - \vet{\ell} \cdot \vet{v}_P \right)
}
{
\lambda_P
+
\left( 1 - \vet{\ell} \cdot \vet{v}_P \right)^2
}
\,.
\label{107}
\end{equation}
Hence,
the value of the parameter $\xi$
that determines the average energy and momentum
of the produced neutrino state
depends on the characteristics of the
production process through the
quantities
$\lambda_P$ and $\vet{v}_P$.

Using Eqs.~(\ref{414}), (\ref{415})
and the relativistic approximations (\ref{101}), (\ref{102}),
with $\xi$ given by Eq.~(\ref{107}),
we find
\begin{equation}
S^P_a(\vet{p}_a)
=
\left(
\frac{m_a^2}{4E\sigma_{pP}}
\right)^2
\left[
\lambda_P
+
\left( 1 - \vet{\ell} \cdot \vet{v}_P \right)^2
\right]^{-1}
\,.
\label{445}
\end{equation}
Hence,
the condition (\ref{411})
is satisfied if
\begin{equation}
m_a^2 \lesssim \sigma_{pP} \, E \,
\sqrt{ \lambda_P + \left( 1 - \vet{\ell} \cdot \vet{v}_P \right)^2 }
\label{419}
\end{equation}
for all values of $a$.
This inequality shows that
a finite momentum uncertainty
$\sigma_{pP}$
of the production process
is necessary in order to produce coherently different
extremely relativistic
massive neutrino components.

The momentum and energy uncertainties
in Eqs.~(\ref{454}) and (\ref{456})
can be estimated at zeroth order in $m_a^2/E^2$
inverting the symmetric matrix
\begin{equation}
\Lambda
=
\frac{1}{2\sigma_{pP}^2}
\begin{pmatrix}
1 + \frac{\left(v_P^x-1\right)^2}{\lambda_P}
&
\frac{\left(v_P^x-1\right)v_P^y}{\lambda_P}
&
\frac{\left(v_P^x-1\right)v_P^z}{\lambda_P}
\\
\frac{\left(v_P^x-1\right)v_P^y}{\lambda_P}
&
1 + \frac{(v_P^y)^2}{\lambda_P}
&
\frac{v_P^y v_P^z}{\lambda_P}
\\
\frac{\left(v_P^x-1\right)v_P^z}{\lambda_P}
&
\frac{v_P^y v_P^z}{\lambda_P}
&
1 + \frac{(v_P^z)^2}{\lambda_P}
\end{pmatrix}
\,.
\label{124}
\end{equation}
From the determinant
\begin{equation}
\mathrm{Det}\Lambda
=
\frac{1}{(2\sigma_{pP}^2)^3}
\left[
1
+
\frac{(v_P^x-1)^2+(v_P^y)^2+(v_P^z)^2}{\lambda_P}
\right]
\,,
\label{457}
\end{equation}
we obtain\footnote{
I would like to thank M. Beuthe for
finding misprints in Eqs.~(\ref{462}) and (\ref{463})
in the first version
of the paper appeared in the electronic archive hep-ph.
}
\begin{align}
\null & \null
\langle (\delta{p^x})^2 \rangle_a
\simeq
\frac{1}{2}
\,
(\Lambda^{-1})^{xx}
=
\sigma_{pP}^2
\,
\frac
{ 1 + \frac{(v_P^y)^2+(v_P^z)^2}{\lambda_P} }
{ 1 + \frac{(v_P^x-1)^2+(v_P^y)^2+(v_P^z)^2}{\lambda_P} }
\,,
\label{461}
\\
\null & \null
\langle (\delta{p^y})^2 \rangle_a
\simeq
\frac{1}{2}
\,
(\Lambda^{-1})^{yy}
=
\sigma_{pP}^2
\,
\frac
{ 1 + \frac{(v_P^x-1)^2+(v_P^z)^2}{\lambda_P} }
{ 1 + \frac{(v_P^x-1)^2+(v_P^y)^2+(v_P^z)^2}{\lambda_P} }
\,,
\label{462}
\\
\null & \null
\langle (\delta{p^z})^2 \rangle_a
\simeq
\frac{1}{2}
\,
(\Lambda^{-1})^{zz}
=
\sigma_{pP}^2
\,
\frac
{ 1 + \frac{(v_P^x-1)^2+(v_P^y)^2}{\lambda_P} }
{ 1 + \frac{(v_P^x-1)^2+(v_P^y)^2+(v_P^z)^2}{\lambda_P} }
\,,
\label{463}
\\
\null & \null
\langle (\delta{E})^2 \rangle_a
\simeq
\frac{1}{2}
\,
(\Lambda^{-1})^{xx}
=
\sigma_{pP}^2
\,
\frac
{ 1 + \frac{(v_P^y)^2+(v_P^z)^2}{\lambda_P} }
{ 1 + \frac{(v_P^x-1)^2+(v_P^y)^2+(v_P^z)^2}{\lambda_P} }
\,.
\label{464}
\end{align}
Thus,
the momentum and energy uncertainties
are of the order of $\sigma_{pP}$,
the total momentum uncertainty in the production process,
unless the factors in Eqs.~(\ref{461})--(\ref{464})
depending on $\lambda_P$ and $\vet{v}_P$
assume extreme values.

In order to illustrate the meaning of our results,
in the following Subsections
we consider as an example
neutrino production in the pion decay process (\ref{pion-decay}) at rest
($\vet{v}_{\pi^+} = 0$)
in four cases.
From Eqs.~(\ref{415}) and (\ref{106}),
the energy and momentum of the neutrino and muon
at zeroth order in $m_a^2/E^2$
are determined by energy-momentum conservation
in the case of massless neutrinos:
\begin{equation}
E
=
p_{\mu^+}
=
\frac{ m_{\pi^+}^2 - m_{\mu^+}^2 }{ 2 m_{\pi^+} }
\simeq
29.8 \, \mathrm{MeV}
\,,
\qquad
E_{\mu^+}
=
\frac{ m_{\pi^+}^2 + m_{\mu^+}^2 }{ 2 m_{\pi^+} }
\simeq
109.8 \, \mathrm{MeV}
\,,
\label{110}
\end{equation}
leading to the muon velocity
\begin{equation}
v_{\mu^+}
=
\frac{ m_{\pi^+}^2 - m_{\mu^+}^2 }{ m_{\pi^+}^2 + m_{\mu^+}^2 }
\simeq
0.27
\,.
\label{111}
\end{equation}
Equation~(\ref{167})
implies that the localized pion at rest
has no energy uncertainty
(at order $\sigma_{p\pi^+}/m_{\pi^+}$).

\subsection{Unlocalized production process}
\label{Prod: Unlocalized}

In the limit
\begin{equation}
\sigma_{p\pi^+} \to 0
\,,
\quad
\sigma_{p\mu^+} \to 0
\quad
\Longrightarrow
\quad
\sigma_{pP} \to 0
\,,
\label{1081}
\end{equation}
the particles taking part
to the production process
and the production process itself are not localized.

In this case
the condition (\ref{419})
is not satisfied
for the coherent production of different extremely relativistic
massive neutrino components
is not valid.
Furthermore,
no deviation from Eq.~(\ref{415})
can satisfy the inequality (\ref{411})
for more than one value of $a$,
because in the limit (\ref{1081})
$S^P_a(\vet{p}_a)$
becomes infinite,
suppressing the production of $\nu_a$
unless
\begin{equation}
\left( \vet{p}_P - \vet{p}_a \right)^2
+
\frac{1}{\lambda_P}
\left[ E_P - E_a
-
\left( \vet{p}_P - \vet{p}_a \right)
\cdot \vet{v}_P \right]^2
=
0
\,.
\label{1082}
\end{equation}
Taking into account the fact that $\lambda_P$ is positive,
Eq.~(\ref{1082})
can be satisfied only if both squares are zero,
\emph{i.e.} if
$\vet{p}_P = \vet{p}_a$
and
$E_P = E_a$
exactly.
Since this constraint can be satisfied only for one value of the index $a$,
only the corresponding massive neutrino is produced.

\subsection{Unlocalized pion}
\label{Prod: Unlocalized pion}

If the momentum of the pion
is determined with high accuracy,
\begin{equation}
\sigma_{p\pi^+} \to 0
\,,
\label{0651}
\end{equation}
the pion is unlocalized,
\begin{equation}
\sigma_{x\pi^+} \to + \infty
\,,
\label{0652}
\end{equation}
and we have
\begin{equation}
\sigma_{xP} = \sigma_{x\mu^+}
\,,
\label{066}
\end{equation}
which imply
(taking into account $\vet{v}_{\pi^+} = 0$)
\begin{equation}
\vet{v}_P = \vet{v}_{\mu^+} = - \vet{\ell} \, v_{\mu^+}
\,,
\qquad
\Sigma_P = v_{\mu^+}^2
\,,
\qquad
\lambda_P = 0
\,.
\label{109}
\end{equation}

From Eq.~(\ref{107}) we get
\begin{equation}
\xi
=
\frac{1}{2}
\left(
1
-
\frac{ m_{\pi^+}^2 }{ m_{\mu^+}^2 }
\right)
\simeq
0.21
\,.
\label{112}
\end{equation}
This is the value of $\xi$
given by exact energy-momentum
conservation in the production process
in the case of pion decay at rest and
different muon momenta for each massive neutrino.

The condition (\ref{419})
in this case reads
\begin{equation}
m_a^2 \lesssim \sigma_{p\mu^+} \, E \left( 1 + v_{\mu^+} \right)
\sim
\sigma_{p\mu^+} \, m_{\pi^+}
\,.
\label{444}
\end{equation}
If the muon propagates in normal matter,
one can estimate $\sigma_{x\mu^+}$
to be given approximately by the inter-atomic distance,
$\sigma_{x\mu^+} \sim 10^{-8} \, \mathrm{cm}$,
which corresponds to
$\sigma_{p\mu^+} \sim 10^{3} \, \mathrm{eV}$.
Thus, the condition (\ref{444}) numerically gives
$m_a^2 \lesssim 10^{11} \, \mathrm{eV}^2$,
that is certainly satisfied.

The production process has a momentum uncertainty
given by Eq.~(\ref{066}),
and an energy uncertainty
$v_{\mu^+}\sigma_{pP}$
due to the muon.
From Eqs.~(\ref{461}) and (\ref{464})
one can see that in this case the massive neutrino components
of the neutrino state (\ref{099}) have no energy uncertainty
and no momentum uncertainty along the direction $\vet{\ell}$ of propagation.
Hence,
in this limiting case the energy and momentum uncertainties
of the massive neutrino components
of the neutrino state
are rather different from the energy and momentum uncertainties
of the production process.
Only the uncertainties of the components of the momentum
orthogonal to the direction of propagation,
given by Eqs.~(\ref{462}) and (\ref{463}),
are equal to the momentum uncertainty $\sigma_{pP}$
of the production process.
These uncertainties are actually necessary in order to localize
the massive neutrino components along the direction of propagation.

\subsection{Equal energy limit}
\label{Prod: Equal energy limit}

In the limit in which the pion is localized
but the final lepton is not localized,
\begin{equation}
\sigma_{x\mu^+} \to + \infty
\,,
\label{113}
\end{equation}
we have
\begin{equation}
\sigma_{xP} = \sigma_{x\pi^+}
\,,
\qquad
\vet{v}_P = 0
\,,
\qquad
\lambda_P = \Sigma_P = 0
\,,
\label{114}
\end{equation}
which imply
\begin{equation}
\xi = 0
\,.
\label{115}
\end{equation}
In this limit the different massive neutrino components
have the same energy.

Since $\sigma_{pP} = \sigma_{p\pi^+} \neq 0$,
the production process has
a momentum uncertainty,
but Eq.~(\ref{167}) implies that both the pion and the muon
have no energy uncertainty.
From Eqs.~(\ref{461}) and (\ref{464})
it follows that
each massive neutrino component in the state (\ref{099})
has definite energy and momentum along the direction of propagation,
without any spread.
This is due to the fact that
the relations (\ref{114})
imply that 
$S^P_a(\vet{p})$ in Eq.~(\ref{SPa}) is infinite,
suppressing the production of $\nu_a$,
unless
$E_{\nu_a}(\vet{p}) = E_P$
exactly.
This constraint can be satisfied simultaneously
for different values of the index $a$ taking different values of $\vet{p}$.
The momentum uncertainty of the pion
is necessary in order to allow the coherent production
of a state with different massive neutrino components
with different values of $\vet{p}$.
For each massive neutrino component
there is only one energy
$E_a=E_P$,
equal for all components,
and the corresponding momentum $p_a = \sqrt{ E_P^2 - m_a^2 }$.
This implies that along the direction of propagation
each massive neutrino component in the state (\ref{099})
is a plane wave and not a wave packet.

Equations~(\ref{462}) and (\ref{463})
imply that
the uncertainties of the components of the momentum
orthogonal to the direction of propagation are equal to
$\sigma_{pP}$.
These uncertainties are allowed because
they generate an energy uncertainty of higher order
in $\sigma_{pP}/E_P$,
that has been neglected in our formalism.
Actually,
as we have already remarked in the discussion of the previous example,
these uncertainties are necessary in order to localize
the massive neutrino components along the direction of propagation.

The case under consideration
has some similarity with
the one considered in Ref.~\cite{Grimus-Stockinger-96}
by Grimus and Stockinger,
in which it was assumed that
the particles that take part in the
production process
(as well as those participating to the detection process)
are in bound states
with definite energy
or are described by plane waves with definite energy.
Although
the physical picture in Ref.~\cite{Grimus-Stockinger-96}
is different from the example under consideration,
in which the localized pion at rest is free,
in both cases
there is no energy uncertainty
in the production process
and
different massive neutrino components
have the same energy
because of exact energy conservation.
In this sense,
the case considered in Ref.~\cite{Grimus-Stockinger-96}
can be considered effectively as a limiting case
of the general wave-packet treatment considered here,
as done in Ref.~\cite{Beuthe:2002ej}
comparing the model of Ref.~\cite{Grimus-Stockinger-96}
with the wave-packet model with virtual intermediate neutrinos
discussed in Refs.~\cite{Giunti-Kim-Lee-Lee-93,Giunti-Kim-Lee-Whendo-98}.

We think that in some cases it may be important
to consider the fact that some particles are in bound states,
but an appropriate description of such a case
should take into account also the fact that
the bound states (as atomic nuclei) are localized,
leading to a wave packet description.
We also think that
the description of some particles taking part to the production
(or detection) process by plane waves
is unrealistic,
because neutrinos are usually produced in dense media,
where these particles are localized by interactions.

In conclusion of this subsection
we would like to remark that,
although the equal energy limit that we have considered is realizable
in principle,
we think that in practice it is rather unlikely
since the pion
(or in general the initial particle $P_I$)
must decay exactly at rest
and the produced muon
(or in general the final particles $P_F$ and $\ell_{\alpha}^{+}$)
must be completely unlocalized.
If the pion does not decay
at rest in the reference frame of the observer,
it is possible to boost the reference frame
to the one in which the pion is at rest,
but the pion velocity in the reference frame of the observer must
be known with high accuracy.
This information is usually not available,
for example in the existing
accelerator and atmospheric neutrino experiments
in which neutrinos are produced by pion decay in flight.

\subsection{Realistic case}
\label{Prod: Realistic case}

In a realistic experimental setup
the localizations of the pion and muon are of the same order of magnitude.
Indeed, the typical neutrino production in pion decay at rest\footnote{
In accelerator experiments in which neutrinos are produced by pion decay in flight
the localization of the pion and muon are given by the dimensions of the decay tunnel.
In solar neutrino experiments
electron neutrinos are produced in the core of the sun
where all the particles taking part to the production process are
localized by interactions with the dense medium.
In atmospheric neutrino experiments
neutrinos are produced by pion and muon decay in flight in the atmosphere,
where pions, muons and electrons are localized by the interactions with air.
In reactor neutrinos experiments electron antineutrinos are produced
by the decays of heavy elements in the dense reactor core
where the heavy nuclei and electrons are localized by
interactions with the medium.
}
occurs in a medium,
where both the pion and muon are localized
by interactions with the surrounding atoms.
Let us consider, for example,
\begin{equation}
\sigma_{x\pi^+}
\simeq
\sigma_{x\mu^+}
\simeq
2 \, \sigma_{xP}
\,,
\label{117}
\end{equation}
which leads to
\begin{equation}
\vet{v}_P
\simeq
\frac{\vet{v}_{\mu^+}}{2}
=
- \frac{v_{\mu^+}}{2} \, \vet{\ell}
\,,
\qquad
\Sigma_P
\simeq
\frac{v_{\mu^+}^2}{2}
\,,
\qquad
\lambda_P
\simeq
\frac{v_{\mu^+}^2}{4}
\,.
\label{118}
\end{equation}
In this case we have
\begin{equation}
\xi
\simeq
\frac
{
\frac{1}{2}
\left(
1
-
\frac{ m_{\pi^+}^2 }{ m_{\mu^+}^2 }
\right)
}
{
1
+
\frac{1}{4}
\left(
1
+
\frac{ m_{\pi^+}^2 }{ m_{\mu^+}^2 }
\right)^2
}
\simeq
0.13
\,.
\label{119}
\end{equation}
Hence we see that the value of $\xi$ in a realistic situation
is of the same order of magnitude as that
in Eq.~(\ref{112}),
corresponding to exact energy-momentum
conservation in the production process
in the case of pion decay at rest and
different muon momenta for each massive neutrino.

The condition (\ref{419})
in this case becomes
\begin{equation}
m_a^2 \lesssim \sigma_{pP} \, E \,
\sqrt{ 1 + v_{\mu^+} + \frac{v_{\mu^+}^2}{2} }
\sim
\sigma_{p\mu^+} \, E
\,.
\label{446}
\end{equation}
If the pion decay occurs in normal matter,
$\sigma_{xP}$
is given approximately by the inter-atomic distance,
$\sigma_{xP} \sim 10^{-8} \, \mathrm{cm}$,
corresponding to
$\sigma_{pP} \sim 10^{3} \, \mathrm{eV}$,
the condition (\ref{446}) numerically reads
$m_a^2 \lesssim 10^{7} \, \mathrm{eV}^2$,
that is certainly satisfied.

Since in this realistic case $\vet{v}_P$ and $\lambda_P$
do not have extreme values,
from Eqs.~(\ref{461})--(\ref{464})
one can see that
the energy and momentum uncertainties of the
neutrino state are of the order of
$\sigma_{pP}$,
about $10^{3} \, \mathrm{eV}$
for pion decay in normal matter.

\section{Detection}
\label{Detection}

Let us consider neutrino detection through the
charged-current weak process
\begin{equation}
\nu_{\beta} + D_I \to D_F + \ell_{\beta}^{-}
\,,
\label{dete}
\end{equation}
at a space-time distance
$(\vet{L},T)$
from the production process.

The state
$| \nu_{\alpha} \rangle$
in Eq.~(\ref{099})
describes the neutrino produced in the process (\ref{prod})
at the origin of the
space-time coordinates.
Since we want to describe with the same formalism
the detection of the neutrino
through the process (\ref{dete})
occurring at a space-time distance
$(\vet{L},T)$
from the production process,
we must translate the origin to the detection space-time point.
Hence,
the neutrino state relevant for the detection process
is obtained by acting on $| \nu_{\alpha} \rangle$
with the space-time translation operator
$\exp\left( -i \widehat{E} T + i \widehat{\vet{P}} \cdot \vet{L} \right)$,
where
$\widehat{E}$ and $\widehat{\vet{P}}$
are the energy and momentum operators,
respectively.
The resulting state is
\begin{equation}
| \nu_{\alpha}(\vet{L},T) \rangle
=
N_{\alpha}
\sum_a U_{{\alpha}a}^{*}
\int
\mathrm{d}^3p
\,
\exp\left( -i E_{\nu_a}(\vet{p}) T + i \vet{p} \cdot \vet{L} \right)
\,
e^{-S^P_a(\vet{p})}
\sum_h
\mathcal{A}^P_a(\vet{p},h)
\,
| \nu_a(\vet{p},h) \rangle
\,.
\label{121}
\end{equation}

The amplitude of interaction of the neutrino state
$| \nu_{\alpha}(\vet{L},T) \rangle$
in the detection process (\ref{dete})
is
\begin{equation}
\mathcal{A}_{\alpha\beta}(\vet{L},T)
=
\langle
D_F , \ell_{\beta}^{-}
|
- i \int \mathrm{d}^4x \,
\mathcal{H}_I^D(x) \,
|
D_I , \nu_{\alpha}(\vet{L},T)
\rangle
\,,
\label{141}
\end{equation}
where
\begin{eqnarray}
\mathcal{H}_I^D(x)
& = &
\frac{G_F}{\sqrt{2}}
\,
\bar\ell_{\beta}(x)
\,
\gamma^{\rho}
\left( 1 - \gamma_5 \right)
\nu_{\beta}(x)
\,
J^{D}_{\rho}(x)
\nonumber
\\
& = &
\frac{G_F}{\sqrt{2}}
\sum_{b}
U_{{\beta}b}
\,
\bar\ell_{\beta}(x)
\,
\gamma^{\rho}
\left( 1 - \gamma_5 \right)
\nu_{b}(x)
\,
J^{D}_{\rho}(x)
\,,
\label{HID}
\end{eqnarray}
is the effective interaction Hamiltonian that
describes the detection process,
and
$J^{D}_{\rho}(x)$
is the weak charged current
that describes the transition
$ D_{I} \to D_{F} $.

For simplicity,
we assume that the particles that take part to the detection process
are described by the Gaussian wave-packet states (\ref{states})
with
$\chi = D_I, D_F, \ell_{\beta}^{-}$.
Using the
neutrino state (\ref{121})
and the Fourier expansion (\ref{fer})
for the lepton fields,
respectively,
we obtain
\begin{align}
\mathcal{A}_{\alpha\beta}(\vet{L},T)
\propto
\null & \null
\sum_a U_{{\alpha}a}^{*} U_{{\beta}a}
\int \mathrm{d}^4x
\sum_h
\int
\mathrm{d}^3p
\,
\mathcal{A}^P_a(\vet{p},h)
\,
e^{-S^P_a(\vet{p})}
\,
\exp\left[
- i E_{\nu_a}(\vet{p}) T + i \vet{p} \cdot \vet{L}
\right]
\nonumber
\\
\null & \null
\times
\int
\mathrm{d}^3p'_{D_F}
\,
\psi_{D_F}^{*}(\vet{p}'_{D_F};\vet{p}_{D_F},\sigma_{pD_F})
\int
\mathrm{d}^3p'_{\ell^{-}_{\beta}}
\,
\psi_{\ell^{-}_{\beta}}^{*}(\vet{p}'_{\ell^{-}_{\beta}};\vet{p}_{\ell^{-}_{\beta}},\sigma_{\ell^{-}_{\beta}})
\nonumber
\\
\null & \null
\times
\int
\mathrm{d}^3p'_{D_I}
\,
\psi_{D_I}(\vet{p}'_{D_I};\vet{p}_{D_I},\sigma_{pD_I})
\nonumber
\\
\null & \null
\times
\bar{u}_{\ell^{-}_{\beta}}(\vet{p}_{\ell^{-}_{\beta}},h_{\ell^{-}_{\beta}})
\gamma^{\rho}
\left( 1 - \gamma_5 \right)
u_{\nu_a}(\vet{p},h)
\,
J^{D}_{\rho}(\vet{p}'_{D_F},h_{D_F};\vet{p}'_{D_I},h_{D_I})
\,
e^{ i ( p'_{D_F} + p'_{\ell^{-}_{\beta}} - p'_{D_I} - \vet{p} ) x }
\,,
\label{142}
\end{align}
where
$
p^{0}
=
E_{\nu_a}(\vet{p})
$,
where
$
p^{\prime0}_{D_I}
=
E_{D_I}(\vet{p}'_{D_I})
$,
$
p^{\prime0}_{D_F}
=
E_{D_F}(\vet{p}'_{D_F})
$,
$
p^{\prime0}_{\ell^{-}_{\beta}}
=
E_{\ell^{-}_{\beta}}(\vet{p}'_{\ell^{-}_{\beta}})
$,
and
$
J^{D}_{\rho}(\vet{p}'_{D_F},h_{D_F};\vet{p}'_{D_I},h_{D_I})
=
\langle D_F(\vet{p}'_{D_F},h_{D_F}) |
\,
J^{D}_{\rho}(0)
\,
| D_I(\vet{p}'_{D_I},h_{D_I}) \rangle
$.
Following the same method as that used for the production process
(see Eq.~(\ref{p04})),
we perform the integrals over
$\mathrm{d}^3p'_{D_F}$,
$\mathrm{d}^3p'_{\ell^{-}_{\beta}}$
and
$\mathrm{d}^3p'_{D_I}$
with a saddle-point approximation
through the approximation (\ref{energy}),
obtaining
\begin{eqnarray}
\mathcal{A}_{\alpha\beta}(\vet{L},T)
&\propto&
\sum_a U_{{\alpha}a}^{*} U_{{\beta}a}
\sum_h
\int
\mathrm{d}^3p
\,
\mathcal{A}^P_a(\vet{p},h)
\mathcal{A}^D_a(\vet{p},h)
\,
e^{-S^P_a(\vet{p})}
\,
\exp\left[
- i E_{\nu_a}(\vet{p}) T + i \vet{p} \cdot \vet{L}
\right]
\nonumber
\\
&& \null
\times
\int
\mathrm{d}^4x
\,
\exp\left[
- i ( E_{\nu_a}(\vet{p}) - E_D ) t
+ i ( \vet{p} - \vet{p}_D ) \vet{x}
-
\frac
{ \vet{x}^2
- 2 \, \vet{v}_D \cdot \vet{x} \, t
+ \Sigma_D \, t^2 }
{ 4 \sigma_{xD}^2 }
\right]
\,,
\label{151}
\end{eqnarray}
with
\begin{eqnarray}
&& \null
E_D
\equiv
E_{D_F}
+
E_{\ell^{-}_{\beta}}
-
E_{D_I}
\,,
\label{ED}
\\
&& \null
\vet{p}_D 
\equiv
\vet{p}_{D_F}
+
\vet{p}_{\ell^{-}_{\beta}}
-
\vet{p}_{D_I}
\,,
\label{pD}
\\
&& \null
\frac{1}{\sigma_{xD}^2}
\equiv
\frac{1}{\sigma_{xD_{I}}^2}
+
\frac{1}{\sigma_{xD_{F}}^2}
+
\frac{1}{\sigma_{x\ell^{-}_{\beta}}^2}
\,,
\label{sxD}
\\
&& \null
\vet{v}_D
\equiv
\sigma_{xD}^2
\left(
\frac{ \vet{v}_{D_I} }{ \sigma_{xD_{I}}^2 }
+
\frac{ \vet{v}_{D_F} }{ \sigma_{xD_{F}}^2 }
+
\frac{ \vet{v}_{\ell^{-}_{\beta}} }{ \sigma_{x\ell^{-}_{\beta}}^2 }
\right)
\,,
\label{vD}
\\
&& \null
\Sigma_D
\equiv
\sigma_{xD}^2
\left(
\frac{ \vet{v}_{D_I}^2 }{ \sigma_{xD_{I}}^2 }
+
\frac{ \vet{v}_{D_F}^2 }{ \sigma_{xD_{F}}^2 }
+
\frac{ \vet{v}_{\ell^{-}_{\beta}}^2 }{ \sigma_{x\ell^{-}_{\beta}}^2 }
\right)
\,,
\label{SD}
\\
&& \null
\mathcal{A}^D_a(\vet{p},h)
\equiv
\bar{u}_{\ell^{-}_{\beta}}(\vet{p}_{\ell^{-}_{\beta}},h_{\ell^{-}_{\beta}})
\,
\gamma^{\rho}
\left( 1 - \gamma_5 \right)
u_{\nu_a}(\vet{p},h)
\,
J^{D}_{\rho}(\vet{p}'_{D_F},h_{D_F};\vet{p}'_{D_I},h_{D_I})
\,.
\label{ADa}
\end{eqnarray}
As expected,
the overall spatial width $\sigma_{xD}$
of the production process is dominated by the smallest
among the spatial widths of the participating particles.
As the corresponding quantities
in the production process,
$|\vet{v}_D|$ and $\Sigma_D$
are limited by
\begin{equation}
0
\leq
|\vet{v}_D|
\leq
1
\,,
\qquad
0
\leq
\Sigma_D
\leq
1
\,.
\label{439}
\end{equation}

The Gaussian integral over $\mathrm{d}^4x$ leads to
\begin{equation}
\mathcal{A}_{\alpha\beta}(\vet{L},T)
\propto
\sum_a U_{{\alpha}a}^{*} U_{{\beta}a}
\sum_h
\int
\mathrm{d}^3p
\,
\mathcal{A}^P_a(\vet{p},h)
\mathcal{A}^D_a(\vet{p},h)
\,
e^{-S_a(\vet{p})}
\,
\exp\left[
- i E_{\nu_a}(\vet{p}) T + i \vet{p} \cdot \vet{L}
\right]
\,,
\label{152}
\end{equation}
where
\begin{equation}
S_a(\vet{p})
=
S^P_a(\vet{p}) + S^D_a(\vet{p})
\,,
\label{153}
\end{equation}
and $S^D_a(\vet{p})$
has the same structure as $S^P_a(\vet{p})$ given in Eq.~(\ref{SPa}),
with the quantities relative to the production process
replaced by the corresponding ones relative to the detection process:
\begin{equation}
S^D_a(\vet{p})
\equiv
\frac
{ \left( \vet{p}_D - \vet{p} \right)^2 }
{ 4 \, \sigma_{pD}^2 }
+
\frac
{ \left[ \left( E_D - E_{\nu_a}(\vet{p})  \right)
-
\left( \vet{p}_D - \vet{p} \right)
\cdot \vet{v}_D \right]^2 }
{ 4 \, \sigma_{pD}^2 \lambda_D }
\,,
\label{SDa}
\end{equation}
with
\begin{equation}
\lambda_D
\equiv
\Sigma_D - \vet{v}_D^2
\,, 
\label{442}
\end{equation}
limited by
\begin{equation}
0
\leq
\lambda_D
\leq
1
\,,
\label{441}
\end{equation}
and the momentum uncertainty $\sigma_{pD}$ defined by
\begin{equation}
\sigma_{xD} \, \sigma_{pD}
=
\frac{ 1 }{ 2 }
\,,
\label{165}
\end{equation}
which
is dominated by the largest of the momentum uncertainties
of
$D_I$,
$D_F$
and
$\ell^{-}_{\beta}$:
\begin{equation}
\sigma_{pD}^2
=
\sigma_{pD_{I}}^2
+
\sigma_{pD_{F}}^2
+
\sigma_{p\ell^{-}_{\beta}}^2
\,.
\label{spD}
\end{equation}

In general
the dominant momentum contribution
to the integration over $\mathrm{d}^3p$
(\ref{152})
does not come from the stationary point of
$S^P_a(\vet{p})$,
but from the stationary point of
$S_a(\vet{p})$,
which takes into account also the
momentum and energy uncertainties of the detection process.
If these uncertainties
are smaller than those of the production process,
the detection process
picks up as dominant contribution to the
flavor-changing amplitude (\ref{152})
a value of the neutrino momentum in the wave packet (\ref{121})
that is significantly
different from $\vet{p}_a$ corresponding to the stationary point of
$S^P_a(\vet{p})$.

We denote by $\vet{k}_a$ the momentum 
corresponding to the stationary point of
$S_a(\vet{p})$,
\begin{equation}
\left.
\frac{ {\partial}S_a }{ {\partial}\vet{p} }
\right|_{\vet{p}=\vet{k}_a}
=
0
\,,
\label{155}
\end{equation}
which gives the dominant contribution
to the transition amplitude (\ref{152}).
The value of $\vet{k}_a$ is given by
\begin{align}
\null & \null
\frac{
\vet{p}_P
-
\vet{k}_a
}
{ \sigma_{pP}^2 }
+
\frac{
E_P - \varepsilon_a
-
\left( \vet{p}_P - \vet{k}_a \right) \cdot \vet{v}_P
}
{ \sigma_{pP}^2 \, \lambda_P }
\left( \vet{u}_a - \vet{v}_P \right)
\nonumber
\\
+
\null & \null
\frac{
\vet{p}_D
-
\vet{k}_a
}
{ \sigma_{pD}^2 }
+
\frac{
E_D - \varepsilon_a
-
\left( \vet{p}_D - \vet{k}_a \right) \cdot \vet{v}_D
}
{ \sigma_{pD}^2 \, \lambda_D }
\left( \vet{u}_a - \vet{v}_D \right)
=
0
\,,
\label{ka}
\end{align}
with
\begin{eqnarray}
&& \null
\varepsilon_a
\equiv
E_{\nu_a}(\vet{k}_a)
=
\sqrt{ \vet{k}_a^2 + m_a^2 }
\,,
\label{epsa}
\\
&& \null
\vet{u}_a
\equiv
\left.
\frac{ {\partial}E_{\nu_a}(\vet{p}) }{ {\partial}\vet{p} }
\right|_{\vet{p}=\vet{k}_a}
=
\frac{\vet{k}_a}{\varepsilon_a}
\,.
\label{ua}
\end{eqnarray}
We have chosen the notation
$\vet{k}_a$,
$\varepsilon_a$
and
$\vet{u}_a$
in order to emphasize that in general these quantities
are not the average momenta, energies and velocities
of the neutrino wave packets propagating
between production and detection,
denoted by
$\vet{p}_a$,
$E_a$
and
$\vet{v}_a$,
which
are determined only by the production process,
as follows from Eqs.~(\ref{pnua})--(\ref{vnua}).
The quantities
$\vet{k}_a$,
$\varepsilon_a$
and
$\vet{u}_a$
are approximately equal
to the average neutrino wave packets
momenta, energies and velocities
$\vet{p}_a$,
$E_a$
and
$\vet{v}_a$
only if
$
\sigma_{pD}^2
\gg
\sigma_{pP}^2
$
and
$
\sigma_{pD}^2 \, \lambda_D
\gg
\sigma_{pP}^2 \, \lambda_P
$.
In other words,
the properties
of the neutrino wave packets
can be measured only with a detection process having
a relatively large
momentum and energy uncertainty,
which correspond to a relatively sharp
spatial and temporal localization.

Before solving Eq.~(\ref{ka}),
we notice that the amplitude (\ref{152})
is not suppressed only if 
\begin{equation}
S_a(\vet{k}_a)
\lesssim
1
\label{421}
\end{equation}
for all values of $a$.
The inequality (\ref{421}),
together with Eq.~(\ref{ka}),
constraint the possible values of
$\vet{k}_a$, $\vet{p}_P$, $E_P$, $\vet{p}_D$ and $E_D$.
Since both $S^P_a(\vet{k}_a)$ and $S^D_a(\vet{k}_a)$
are positive,
we have the conditions
$
S^P_a(\vet{k}_a) \ll 1
$
and
$
S^D_a(\vet{k}_a) \ll 1
$
for all values of $a$.
The first inequality is similar to the condition (\ref{411}),
but now it concerns the momenta $\vet{k}_a$ and energies $\varepsilon_a$
that are relevant for the flavor transition amplitude (\ref{152}).

Similarly to what we have done in the discussion of the production process
(see Eq.~(\ref{415})),
we consider
\begin{equation}
\vet{p}_P = E_P \, \vet{\ell}
\,,
\qquad
\vet{p}_D = E_D \, \vet{\ell}
\,,
\qquad
E_P = E_D
\,,
\label{425}
\end{equation}
which
corresponds to exact energy and momentum conservation
in the production and detection processes
in the case of massless neutrinos.
Here
$\vet{\ell}\equiv\vet{L}/|\vet{L}|$
is the unit vector in the direction of propagation of the neutrino
from the production to the detection processes.
It is possible to consider deviations
from the relations (\ref{425}) compatible with the inequality
(\ref{421}),
but such deviations would introduce
many complications in the following formalism,
without further insights in the physics of neutrino oscillations.

Equation~(\ref{ka}) implies that the massive neutrino momenta
$\vet{k}_a$ must be aligned in the $\vet{\ell}$ direction:
\begin{equation}
\vet{k}_a = k_a \, \vet{\ell}
\,.
\label{424}
\end{equation}
We solve Eq.~(\ref{ka})
in the relativistic approximation
\begin{align}
\null & \null
\varepsilon_a \simeq E + \rho \, \frac{m_a^2}{2E}
\,,
\label{158}
\\
\null & \null
k_a \simeq E - \left( 1 - \rho \right) \frac{m_a^2}{2E}
\,.
\label{159}
\end{align}
At zeroth order in $m_a^2/E^2$ we obtain
\begin{equation}
E = E_P = E_D
\,,
\label{160}
\end{equation}
and at first order
\begin{equation}
\rho
=
\frac
{
\frac{1}{\sigma_{p}^2}
-
\frac
{ \vet{\ell} \cdot \vet{v}_P \left( 1 - \vet{\ell} \cdot \vet{v}_P \right) }
{ \sigma_{pP}^2 \, \lambda_P }
-
\frac
{ \vet{\ell} \cdot \vet{v}_D \left( 1 - \vet{\ell} \cdot \vet{v}_D \right) }
{ \sigma_{pD}^2 \, \lambda_D }
}
{
\frac{1}{\sigma_{p}^2}
+
\frac
{ \left( 1 - \vet{\ell} \cdot \vet{v}_P \right)^2 }
{ \sigma_{pP}^2 \, \lambda_P }
+
\frac
{ \left( \vet{\ell} \cdot \vet{v}_D - 1 \right)^2 }
{ \sigma_{pD}^2 \, \lambda_D }
}
\,,
\label{161}
\end{equation}
with
\begin{equation}
\frac{1}{\sigma_{p}^2}
=
\frac{1}{\sigma_{pP}^2}
+
\frac{1}{\sigma_{pD}^2}
\,.
\label{162}
\end{equation}
Notice that in the production and detection processes
the squared momentum uncertainties add,
as shown in Eqs.~(\ref{sxP}) and (\ref{sxD}),
whereas
in the total amplitude the inverses of the squared momentum uncertainties
of the production and detection processes add.
This is expected on the basis of simple physical arguments.
Indeed,
a large momentum uncertainty of a particle must increase the total
momentum uncertainty in the corresponding process,
whereas
a small momentum uncertainty in one of the two processes
constraints the momentum uncertainty of the neutrino
connecting the two processes leading to a restriction
of the momentum uncertainty in the other process.
On the other hand,
in the production and detection processes
the inverses of the squared space uncertainties add
(see Eqs.~(\ref{sxP}) and (\ref{sxD})),
whereas
in the total amplitude the squared space uncertainties
of the production and detection processes add:
\begin{equation}
\sigma_{x}^2
=
\sigma_{xP}^2
+
\sigma_{xD}^2
\,,
\label{1621}
\end{equation}
with $\sigma_{x}$ defined by the relation
\begin{equation}
\sigma_{x} \, \sigma_{p} = \frac{1}{2}
\,.
\label{1622}
\end{equation}
Also the behavior in Eq.~(\ref{1621})
is expected on the basis of simple physical arguments:
a small space uncertainty of a particle
localizes better the corresponding process,
whereas a large space uncertainty of
one of the two processes
increases the coherence of the overall process.

In the relativistic approximation
$S_a(\vet{k}_a)$
is given by
\begin{equation}
S_a(\vet{k}_a)
=
\zeta
\left(
\frac{m_a^2}{4E\sigma_{p}}
\right)^2
\,,
\label{471}
\end{equation}
with
\begin{equation}
\zeta
=
\frac
{
\frac{1}{\sigma_{p}^2}
\left(
\frac{1}{\sigma_{pP}^2\lambda_P}
+
\frac{1}{\sigma_{pD}^2\lambda_D}
\right)
+
\left(
\frac{1 - \vet{\ell} \cdot \vet{v}_P}{\sigma_{pP}^2\lambda_P}
+
\frac{1 - \vet{\ell} \cdot \vet{v}_D}{\sigma_{pD}^2\lambda_D}
\right)^2
+
\frac
{\left[ \vet{\ell} \cdot \left( \vet{v}_P - \vet{v}_D \right) \right]^2}
{\sigma_{pP}^2\lambda_P \, \sigma_{pD}^2\lambda_D}
\left(
2
+
\frac{\sigma_{p}^2}{\sigma_{pP}^2}
\,
\frac{\left( 1 - \vet{\ell} \cdot \vet{v}_P \right)^2}{\lambda_P}
+
\frac{\sigma_{p}^2}{\sigma_{pD}^2}
\,
\frac{\left( 1 - \vet{\ell} \cdot \vet{v}_D \right)^2}{\lambda_D}
\right)
}
{
\left[
\frac{1}{\sigma_{p}^2}
+
\frac{\left( 1 - \vet{\ell} \cdot \vet{v}_P \right)^2}{\sigma_{pP}^2\lambda_P}
+
\frac{\left( 1 - \vet{\ell} \cdot \vet{v}_D \right)^2}{\sigma_{pD}^2\lambda_D}
\right]^2
}
\,.
\label{472}
\end{equation}
It follows that
in the relativistic approximation
the condition (\ref{421}) reads
\begin{equation}
m_a^2 \lesssim \sigma_{p} \, E \, \zeta^{-1/2}
\,.
\label{473}
\end{equation}
If this inequality is satisfied for all values of the index $a$,
the choices (\ref{425}) are acceptable
and the different extremely relativistic
massive neutrino components contribute
coherently to the flavor-changing
transition amplitude.

Let us return to the calculation of the transition amplitude.
The integral over $\mathrm{d}^3p$ in Eq.~(\ref{152})
can be performed with a saddle-point approximation around the stationary
point $\vet{k}_a$ of $S_a(\vet{p})$,
leading to
\begin{eqnarray}
\mathcal{A}_{\alpha\beta}(\vet{L},T)
&\propto&
\sum_a U_{{\alpha}a}^{*} U_{{\beta}a}
\sum_h
\frac{
\mathcal{A}^P_a(\vet{k}_a,h)
\mathcal{A}^D_a(\vet{k}_a,h)
}
{ \sqrt{ \mathrm{Det}\Omega_a } }
\,
e^{-S_a(\vet{k}_a)}
\nonumber
\\
&&
\times
\exp\left[
- i \varepsilon_a T + i \vet{k}_a \cdot \vet{L}
- \frac{1}{2}
\left( L^j - u_a^j T \right)
(\Omega_a^{-1})^{jk}
\left( L^k - u_a^k T \right)
\right]
\,,
\label{168}
\end{eqnarray}
where
\begin{eqnarray}
\Omega_a^{jk}
&\equiv&
\left.
\frac{ \partial^2 S_a(\vet{p}) }{ \partial p^j \partial p^k }
\right|_{\vet{p}=\vet{k}_a}
\nonumber
\\
&=&
\frac{\delta^{jk}}{2\sigma_{p}^2}
+
\frac{\left(v_P^j-u_a^j\right)\left(v_P^k-u_a^k\right)}{2\sigma_{pP}^2\lambda_P}
-
\frac
{
\left[
\left( E_P - \varepsilon_a  \right)
-
\left( \vet{p}_P - \vet{k}_a \right) \cdot \vet{v}_P
\right]
}
{2\sigma_{pP}^2\lambda_P}
\,
\frac{\delta^{jk} - u_a^j u_a^k}{\varepsilon_a}
\nonumber
\\
&&
\phantom{
\frac{\delta^{jk}}{2\sigma_{p}^2}
}
+
\frac{\left(v_D^j-u_a^j\right)\left(v_D^k-u_a^k\right)}{2\sigma_{pD}^2\lambda_D}
-
\frac
{
\left[
\left( E_D - \varepsilon_a  \right)
-
\left( \vet{p}_D - \vet{k}_a \right) \cdot \vet{v}_D
\right]
}
{2\sigma_{pD}^2\lambda_D}
\,
\frac{\delta^{jk} - u_a^j u_a^k}{\varepsilon_a}
\,.
\label{169}
\end{eqnarray}

The factors
\begin{equation}
- \frac{1}{2}
\left( L^j - u_a^j T \right)
(\Omega_a^{-1})^{jk}
\left( L^k - u_a^k T \right)
\label{126}
\end{equation}
in the exponential of Eq.~(\ref{168})
do not suppress the transition amplitude only if
$\vet{L}$ is aligned with the
common direction $\vet{\ell}$ of the momenta $\vet{k}_a$,
apart from a deviation of the order of
$\Omega_a/L$,
which is very small if the localizations of the
production and detection processes are much smaller
than the source-detector distance,
a condition which is necessary for the observation of neutrino oscillations
and which is satisfied in all neutrino oscillation experiments.
Therefore, we consider
$\vet{L}=L\vet{\ell}$
and
we align $\vet{\ell}$ along the direction of the $x$ axis,
Thus,
the amplitude (\ref{168})
can be written as
\begin{equation}
\mathcal{A}_{\alpha\beta}(\vet{L},T)
\propto
\sum_a U_{{\alpha}a}^{*} U_{{\beta}a}
\sum_h
\frac{
\mathcal{A}^P_a(\vet{k}_a,h)
\mathcal{A}^D_a(\vet{k}_a,h)
}
{ \sqrt{ \mathrm{Det}\Omega_a } }
\,
e^{-S_a(\vet{k}_a)}
\,
\exp\left[
- i \varepsilon_a T + i k_a L
- \frac{ \left( L - u_a T \right)^2 }{ 4 \eta_a^2 }
\right]
\,,
\label{170}
\end{equation}
where
\begin{equation}
\eta_a
\equiv
\sqrt{ \frac{1}{2(\Omega_a^{-1})^{xx}} }
\label{171}
\end{equation}
are the spatial coherence widths.

In the relativistic approximation
the product
$
\mathcal{A}^P_a(\vet{k}_a,+)
\mathcal{A}^D_a(\vet{k}_a,+)
$
corresponding to the
positive helicity component of the massive neutrino $\nu_a$
is suppressed by the ratio $m_a^2/E^2$
with respect to the product
$
\mathcal{A}^P_a(\vet{k}_a,-)
\mathcal{A}^D_a(\vet{k}_a,-)
$
corresponding to the
negative helicity component
and can be neglected.
In the same approximation,
the factors
\begin{equation}
\frac{
\mathcal{A}^P_a(\vet{k}_a,-)
\mathcal{A}^D_a(\vet{k}_a,-)
}
{ \sqrt{ \mathrm{Det}\Omega_a } }
\,
e^{-S_a(\vet{k}_a)}
\,,
\label{172}
\end{equation}
can be approximated with their value
in the case of massless neutrinos,
can be
extracted out of the sum over the
index $a$
and absorbed in the overall normalization factor
of the flavor transition amplitude.
The factorization of these quantities allows
the calculation of the flavor transition amplitude
independently
from the production and detection rates.
The oscillation probability
obtained from this flavor transition amplitude
can be used in the usual calculations
of event rates in neutrino oscillation experiments,
given by the product
of the neutrino flux
calculated for massless neutrinos,
times the oscillation probability,
times the detection cross section
calculated for massless neutrinos
(see \cite{Bilenky-Pontecorvo-PR-78,%
Bilenky-Petcov-RMP-87,%
CWKim-book-93,%
BGG-review-98}).

It is important to notice,
however,
that
a special care is needed for the factors
$e^{-S_a(\vet{k}_a)}$,
to make sure that they do not suppress
the contribution of the corresponding massive neutrino component.
The factors
$e^{-S_a(\vet{k}_a)}$
can be can be approximated with their value
in the case of massless neutrinos only if
\begin{equation}
S_a(\vet{k}_a) \ll 1
\label{1751}
\end{equation}
for all values of the index $a$.
Hence,
the condition (\ref{473}) must be strengthened to
\begin{equation}
m_a^2 \ll \sigma_{p} \, E \, \zeta^{-1/2}
\label{1752}
\end{equation}
for all values of $a$.
Let us emphasize that this condition
is necessary for the unsuppressed
production and detection
of the different extremely relativistic massive neutrino components
whose interference generates neutrino oscillations.
In principle,
it is possible to consider degenerate neutrinos
for which the massless approximation is not appropriate,
but as noted in footnote~\ref{degenerate},
this case is irrelevant in practice.
In any case,
considering a case in which the massless approximation is not appropriate
would require the inclusion of the neutrino mass effect,
which are normally neglected,
in the calculation of the production and detection rates.

If the condition (\ref{1752})
is satisfied
for all values of $a$,
the transition amplitude in the relativistic approximation
can be written as
\begin{equation}
\mathcal{A}_{\alpha\beta}(\vet{L},T)
\propto
\sum_a U_{{\alpha}a}^{*} U_{{\beta}a}
\exp\left[
- i \varepsilon_a T + i k_a L
- \frac{ \left( L - u_a T \right)^2 }{ 4 \eta^2 }
\right]
\,.
\label{175}
\end{equation}
Here $\varepsilon_a$ and $k_a$
are given by their relativistic approximations (\ref{158}) and (\ref{159}),
and
\begin{equation}
u_a
\simeq
1 - \frac{m_a^2}{2E^2}
\,.
\label{3071}
\end{equation}
The coherence widths
$\eta_a$
have been approximated with their value
$\eta$
at zeroth order in $m_a^2/E^2$.

The value of $\eta$
is given by the inversion of the
zeroth order approximation in $m_a^2/E^2$
of the symmetric matrix
$\Omega_a$ in Eq.~(\ref{169}),
\begin{equation}
\Omega_a^{jk}
=
\frac{\delta^{jk}}{2\sigma_{p}^2}
+
\frac{\left(v_P^j-\delta^{jx}\right)\left(v_P^k-\delta^{kx}\right)}{2\sigma_{pP}^2\lambda_P}
+
\frac{\left(v_D^j-\delta^{jx}\right)\left(v_D^k-\delta^{kx}\right)}{2\sigma_{pD}^2\lambda_D}
+
\mathrm{O}\left(\frac{m_a^2}{E^2}\right)
\,.
\label{173}
\end{equation}
We obtained
\begin{equation}
\eta^2
=
\omega
\,
\sigma_{x}^2
\,,
\label{177}
\end{equation}
with
\begin{align}
\omega
=
\null & \null
\left\{
1
+
\sigma_{p}^2
\left[
\frac{\left(v_P^x-1\right)^2+(v_P^y)^2+(v_P^z)^2}{\sigma_{pP}^2\lambda_P}
+
\frac{\left(v_D^x-1\right)^2+(v_D^y)^2+(v_D^z)^2}{\sigma_{pD}^2\lambda_D}
\right]
\right.
\nonumber
\\
\null & \null
\left.
+
\sigma_{p}^4
\,
\frac
{
\left[
\left(v_P^x-1\right) v_D^y
-
\left(v_D^x-1\right) v_P^y
\right]^2
+
\left[
\left(v_P^x-1\right) v_D^z
-
\left(v_D^x-1\right) v_P^z
\right]^2
+
\left(v_P^y v_D^z - v_P^z v_D^y\right)^2
}
{ \sigma_{pP}^2\lambda_P \, \sigma_{pD}^2\lambda_D }
\right\}
\nonumber
\\
\null & \null
\times
\left\{
1
+
\sigma_{p}^2
\left[
\frac{(v_P^y)^2+(v_P^z)^2}{\sigma_{pP}^2\lambda_P}
+
\frac{(v_D^y)^2+(v_D^z)^2}{\sigma_{pD}^2\lambda_D}
\right]
+
\sigma_{p}^4
\,
\frac
{ \left(v_P^y v_D^z - v_P^z v_D^y\right)^2 }
{ \sigma_{pP}^2\lambda_P \, \sigma_{pD}^2\lambda_D }
\right\}^{-1}
\,.
\label{174}
\end{align}

Let us notice that the width $\eta$ obtained in Eq.~(\ref{177})
is somewhat different from the corresponding width obtained
in Ref.~\cite{Giunti-Kim-Lee-Whendo-98}.
The differences stem from the fact that $\Omega_a$ in Eq.~(\ref{169})
is a matrix,
because of the integration over $\mathrm{d}^3p$ in Eq.~(\ref{152}),
whereas the width in Eq.~(18) of Ref.~\cite{Giunti-Kim-Lee-Whendo-98}
is a number obtained from the integration over $\mathrm{d}q^0$ in Eq.~(15)
of Ref.~\cite{Giunti-Kim-Lee-Whendo-98}.

In the limit
$\sigma_{pP} \ll \sigma_{pD}$
the total momentum uncertainty
$\sigma_{p}$
and the coherence width $\eta$
are dominated by the production process, and
$\rho \simeq \xi$,
with $\xi$ given in Eq.~(\ref{107}).
This happens if the production process is much less localized than the
detection process.
In this case
$\vet{k}_a \simeq \vet{p}_a$,
the average momentum of the massive neutrino component $\nu_a$
of the state $|\nu_{\alpha}\rangle$
created in the production process.
In other words,
if the detection process is
much more localized than the production process,
\textit{i.e.} if the momentum uncertainty
of the detection process is much larger than that of the production process,
the transition amplitude
(\ref{175})
depends only on the properties of the neutrino
created in the production process.

However,
in general,
$\sigma_{p}$,
$\eta$
and
$\rho$
depend on both the production and detection processes
and in the limit
$\sigma_{pD} \ll \sigma_{pP}$,
in which the detection process is much less localized than the production process,
they are dominated by the detection process.

The present derivation of
the transition amplitude
in neutrino oscillation experiments
is more complete than the simple quantum-mechanical
model presented in
Refs.~\cite{Giunti-Kim-Lee-Whendo-91,Giunti-Kim-Coherence-98},
in which the energies and momenta of the massive neutrino components
contributing to the transition amplitude
are undetermined and have to be assumed.

Formally,
the transition amplitude (\ref{152})
can be derived by projecting the state
$| \nu_{\alpha}(\vet{L},T) \rangle$
in Eq.~(\ref{121})
on the state
\begin{equation}
| \nu_{\beta} \rangle
=
N_{\beta}
\sum_a U_{{\beta}a}^{*}
\int
\mathrm{d}^3p
\,
e^{-S^D_a(\vet{p})}
\sum_h
\mathcal{A}^D_a(\vet{p},h)
\,
| \nu_a(\vet{p},h) \rangle
\,,
\label{481}
\end{equation}
representing the detection process:
\begin{equation}
\mathcal{A}_{\alpha\beta}(\vet{L},T)
\propto
\langle \nu_\beta | \nu_{\alpha}(\vet{L},T) \rangle
\,,
\label{482}
\end{equation}
as in the quantum-mechanical derivation of
the flavor transition probability in neutrino oscillation experiments.
However,
the calculation of the
coefficients of the massive neutrino components
in the states
$| \nu_{\alpha}(\vet{L},T) \rangle$
and
$| \nu_{\beta} \rangle$
require the quantum field theoretical framework that we
have adopted.

\section{Transition Probability}
\label{Probability}

In this Section we calculate and discuss the transition probability in space
obtained from the average of
\begin{equation}
P_{\alpha\beta}(\vet{L},T)
\propto
|\mathcal{A}_{\alpha\beta}(\vet{L},T)|^2
\label{1781}
\end{equation}
over the unmeasured propagation time $T$,
as done in Refs.~\cite{Giunti-Kim-Lee-Whendo-91,Giunti-Kim-Coherence-98,
Giunti-Kim-Lee-Lee-93,Giunti-Kim-Lee-Whendo-98}.
Let us notice,
however,
that although in present neutrino oscillation experiments
only the source-detector distance $L$ is known,
in the future it may be possible to measure also the
propagation time $T$,
leading to the experimental relevance of
the space-time dependent transition probability (\ref{1781})
(see Refs.~\cite{Anikeev:1998sy,Okun:2000-SNEGIRI}).

From the Gaussian integration over $\mathrm{d}T$ of
$P_{\alpha\beta}(\vet{L},T)$,
in the relativistic approximation
we obtain\footnote{
We order the massive neutrinos by increasing mass:
$m_1 \leq m_2 \leq \ldots$.
In this way the $\Delta{m}^2_{ab}$'s
in the sum over $a>b$ are all positive.
}
\begin{equation}
P_{\alpha\beta}(\vet{L})
=
\sum_a |U_{{\alpha}a}|^2 |U_{{\beta}a}|^2
+
2 \mathrm{Re}
\sum_{a>b} U_{{\alpha}a}^{*} U_{{\beta}a} U_{{\alpha}b} U_{{\beta}b}^{*}
\exp\left[
- 2 \pi i \, \frac{L}{L^{\mathrm{osc}}_{ab}}
- \left( \frac{L}{L^{\mathrm{coh}}_{ab}} \right)^2
- 2 \pi^2 \rho^2 \omega \left( \frac{\sigma_{x}}{L^{\mathrm{osc}}_{ab}} \right)^2
\right]
\,,
\label{178}
\end{equation}
where
\begin{equation}
L^{\mathrm{osc}}_{ab}
=
\frac{4 \pi E}{|\Delta{m}^2_{ab}|}
\label{179}
\end{equation}
are the oscillation lengths,
and
\begin{equation}
L^{\mathrm{coh}}_{ab}
=
\frac{4 \sqrt{2\omega} E^2}{|\Delta{m}^2_{ab}|}
\,
\sigma_{x}
\label{180}
\end{equation}
are the coherence lengths.

The transition probability (\ref{178})
has the same form as that obtained
in Ref.~\cite{Giunti-Kim-Lee-Whendo-98}
in quantum field theory with virtual propagating neutrinos.
The value of $\rho$
in quantum field theory with virtual propagating neutrinos,
derived in Ref.~\cite{Beuthe:2002ej},
is the same as the one derived here (see Eq.~(\ref{161})).
Instead,
the value of $\omega$
is somewhat different from that derived in Ref.~\cite{Giunti-Kim-Lee-Whendo-98},
as already remarked after Eq.~(\ref{174}).

The last term in the exponential of the transition probability
(\ref{178})
suppresses the corresponding oscillatory term
unless the localization of the production and detection processes
is much smaller than the oscillation length,
\begin{equation}
\sigma_x \ll L^{\mathrm{osc}}_{ab}
\,.
\label{483}
\end{equation}
This is due
to the average of the transition probability
(\ref{1781})
over the unmeasured propagation time $T$.
Indeed,
a spatial uncertainty
$\sigma_x$
imply a similar time uncertainty,
as one can understand from the
relation between the energy and momentum uncertainties
in Eq.~(\ref{167}),
or by noticing that
the spatial region in which a process can occur coherently
must be causally connected.
If the time uncertainty,
of the order of $\sigma_x$,
is larger than the oscillation length,
the average of an oscillatory term over the propagation time $T$
is an average over all oscillation phases which depend on $T$,
and the result is zero.

However,
as discussed in Ref.~\cite{Beuthe:2002ej},
the condition (\ref{483})
for the observability of neutrino oscillations
is not necessary if
\begin{equation}
\rho^2 \, \omega = 0
\,.
\label{502}
\end{equation}
As shown in the example
presented in Subsection~\ref{Dete: Equal energy limit},
this can be obtained with
$\rho=0$.
In this case,
since
the different massive neutrino components
have the same energy
$\varepsilon_a = E$,
the oscillation phases do not depend on $T$
and
the average over $T$
of the space-time dependent transition probability (\ref{1781})
has no effect.
Furthermore,
in this case
the coherence lengths
$L^{\mathrm{coh}}_{ab}$
in Eq.~(\ref{180})
can be increased without limit
\cite{Beuthe:2002ej}.
However,
one must notice that this unlimited increase
must be obtained by increasing $\omega$
and not
$\sigma_{x}$,
because an unlimited increase of $\sigma_{x}$
would bring $\sigma_{p}\to0$
and a violation of the condition (\ref{1752})
necessary for the coherent production
of the different massive neutrino components
whose interference generates the oscillations.

In the following Subsections
we consider three cases analogous to those
considered in Subsections~\ref{Prod: Unlocalized},
\ref{Prod: Equal energy limit}
and
\ref{Prod: Realistic case},
that illustrate the effects
of the detection process.
In all these cases we consider the
initial detection particle
$D_I$
at rest
($\vet{v}_{D_I}=0$).

\subsection{Unlocalized detection process}
\label{Dete: Unlocalized}

If the particles taking part to the detection process
are unlocalized,
we have the limit
\begin{equation}
\sigma_{pD_I} \to 0
\,,
\quad
\sigma_{pD_F} \to 0
\,,
\quad
\sigma_{p\ell_{\beta}^{-}} \to 0
\quad
\Longrightarrow
\quad
\sigma_{pD} \to 0
\quad
\Longrightarrow
\quad
\sigma_{p} \to 0
\,.
\label{163}
\end{equation}
In this case the condition (\ref{473})
is not satisfied.
Moreover,
similarly to the case of an unlocalized production process
discussed in Subsection~\ref{Prod: Unlocalized},
no deviation from Eq.~(\ref{425})
can satisfy the inequality (\ref{421})
for more than one value of $a$.
Indeed,
in the limit (\ref{163})
$S^D_a(\vet{k}_a)$
becomes infinite and suppresses the production of $\nu_a$
unless
\begin{equation}
\left( \vet{p}_D - \vet{k}_a \right)^2
+
\frac{1}{\lambda_D}
\left[ E_D - \varepsilon_a
-
\left( \vet{p}_D - \vet{k}_a \right)
\cdot \vet{v}_D \right]^2
=
0
\,.
\label{1631}
\end{equation}
Since $\lambda_D$ is positive,
Eq.~(\ref{1631})
can be satisfied only if both squares are zero,
\emph{i.e.} if
$\vet{p}_D = \vet{k}_a$
and
$E_D = \epsilon_a$
exactly.
Since this constraint can be satisfied only for one value of the index $a$,
only the corresponding massive neutrino is detected
and there are no oscillations,
which are due to the interference
of different massive neutrino components.

\subsection{Equal energy limit}
\label{Dete: Equal energy limit}

Let us consider a localized initial detection particle
$D_I$
at rest,
and
unlocalized final detection particles,
\begin{equation}
\sigma_{pD_F} = \sigma_{p\ell^{-}_{\beta}} = 0
\,,
\label{266}
\end{equation}
which imply
\begin{equation}
\sigma_{pD} = \sigma_{pD_I}
\,,
\qquad
\vet{v}_D = 0
\,,
\qquad
\lambda_D
=
\Sigma_D
=
0
\,.
\label{267}
\end{equation}
In this case,
Eq.~(\ref{167}) implies that all the detection particles
have no energy uncertainty.

Equation~(\ref{161}) gives
\begin{equation}
\rho = 0
\,.
\label{268}
\end{equation}
Hence,
the detection process picks up
the momenta $\vet{k}_a$ of the
different massive neutrino components of the state
$| \nu_{\alpha}(\vet{L},T) \rangle$
in Eq.~(\ref{121})
corresponding to the same energy $\varepsilon_a=E$,
corresponding to exact energy conservation
in the detection process.
These momenta are different from the average momenta
$\vet{p}_a$
of the massive neutrino wave packets
that constitute the state
$| \nu_{\alpha}(\vet{L},T) \rangle$,
which are given by Eq.~(\ref{pnua}),
unless the conditions of Subsection~\ref{Prod: Equal energy limit}
are satisfied by the production process.

From Eq.~(\ref{472}) we find
\begin{equation}
\zeta
=
1
+
\frac{\sigma_{p}^2}{\sigma_{pP}^2}
\frac{\left( \vet{\ell} \cdot \vet{v}_P \right)^2}{\lambda_P}
\,.
\label{269}
\end{equation}
Thus,
the condition (\ref{1752})
can be satisfied by a sufficiently large $\sigma_{p}$.

Since the values in Eq.~(\ref{267})
imply that
\begin{equation}
\omega \to + \infty
\,,
\label{202}
\end{equation}
in the case under consideration
\begin{equation}
L^{\mathrm{coh}}_{ab} \to + \infty
\,,
\label{203}
\end{equation}
\textit{i.e.}
the oscillations remain coherent at arbitrarily large distances.
In agreement with the arguments presented in
Ref.~\cite{Beuthe:2002ej},
the last term in the exponential of Eq.~(\ref{178})
does not suppress the oscillations because
\begin{equation}
\rho^2 \, \omega = 0
\,.
\label{204}
\end{equation}

Physically,
the infinity of the coherence lengths is due
to the fact that
the detection process picks up
the momentum components with equal energy of the
different massive neutrino components.
In this case
there are no oscillations in time
and the average over time
of the space-time dependent transition probability
is irrelevant.

Notice, however,
that the coherence lengths are infinite because of the infinity
of $\omega$,
whereas the uncertainty $\sigma_{x}$
has to be finite in order to satisfy the condition
(\ref{1752})
necessary for the coherent production and detection
of the different massive neutrino components.

As we have remarked
in Subsection~\ref{Prod: Equal energy limit}
for the similar limit in the case
of the production process,
the required conditions
($\vet{v}_{D_I}=0$
and Eq.~(\ref{266}))
are unlikely to be achieved
in any realistic experiment.
We think that
in the case of the detection process
the required conditions are even more unlikely
to be achieved than in the case
of the production process,
because neutrino detection
requires in practice to reveal
at least one of the final particles in the detection process.
This implies that these particles
cannot be unlocalized and the condition
(\ref{266})
is unrealistic.

\subsection{Realistic case}
\label{Dete: Realistic case}

In a realistic case
production and detection occur in matter
and
all the detection particles have uncertainties
of the same order of magnitude,
as well as all the production particles,
as we have discussed in Subsection~\ref{Prod: Realistic case}.
Let us consider, for example, the case in which
the order of magnitude of the spatial localization of
the detection particles is much larger than
that of the production particles:
\begin{equation}
\sigma_{xD_I}
\simeq
\sigma_{xD_F}
\simeq
\sigma_{x\ell_{\beta}^{-}}
\simeq
3
\,
\sigma_{xD}
\gg
3
\,
\sigma_{xP}
\simeq
\sigma_{xP_I}
\simeq
\sigma_{xP_F}
\simeq
\sigma_{x\ell_{\alpha}^{+}}
\,,
\label{401}
\end{equation}
which imply
that the momentum uncertainty of the detection process
is much smaller than that of the production process,
and
\begin{equation}
\vet{v}_{D}
\simeq
\frac{1}{3}
\left(
\vet{v}_{D_I}
+
\vet{v}_{D_F}
+
\vet{v}_{\ell_{\beta}^{-}}
\right)
\,,
\quad
\Sigma_{D}
\simeq
\frac{1}{3}
\left(
\vet{v}_{D_I}^2
+
\vet{v}_{D_F}^2
+
\vet{v}_{\ell_{\beta}^{-}}^2
\right)
\,.
\label{402}
\end{equation}

Let us consider,
for example,
a muon neutrino ($\alpha=\mu$)
produced in the pion decay process (\ref{pion-decay}) at rest
and detected as an electron neutrino
($\beta=e$)
in the process
\begin{equation}
\nu_e + {}^{12}\mathrm{C}
\to
{}^{12}\mathrm{N}_{\mathrm{g.s.}}
+
e^-
\,,
\label{403}
\end{equation}
where
${}^{12}\mathrm{N}_{\mathrm{g.s.}}$
is the ground state of ${}^{12}\mathrm{N}$,
with an atomic mass excess
$\Delta{M}({}^{12}\mathrm{N}) \simeq 17.3 \, \mathrm{MeV}$.
Neglecting the nuclear recoil energy,
the energy of the electron is given by
\begin{equation}
E_{e^-}
\simeq
E - \Delta{M}({}^{12}\mathrm{N})
\simeq
12.5 \, \mathrm{MeV}
\,,
\label{404}
\end{equation}
where $E$ is the energy of a massless neutrino given in Eq.~(\ref{110}).
Let us consider a recoil electron
emitted in the backward direction:
\begin{equation}
\vet{v}_{e^-} = - \vet{\ell} \, |\vet{v}_{e^-}|
\,.
\label{405}
\end{equation}
Since
$|\vet{v}_{D_F}| \simeq |\vet{v}_{D_I}| = 0$
and
$|\vet{v}_{e^-}| \simeq 1$,
we have
\begin{equation}
\vet{v}_{D}
\simeq
-
\frac{1}{3}
\,
\vet{\ell}
\,,
\qquad
\Sigma_{D}
\simeq
\frac{1}{3}
\,,
\qquad
\lambda_{D}
\simeq
\frac{1}{9}
\,.
\label{406}
\end{equation}
From Eq.~(\ref{161}) we get
\begin{equation}
\rho
\simeq
\frac
{
\lambda_D
-
\vet{\ell} \cdot \vet{v}_D \left( 1 - \vet{\ell} \cdot \vet{v}_D \right)
}
{
\lambda_D
+
\left( \vet{\ell} \cdot \vet{v}_D - 1 \right)^2
}
\simeq
0.29
\,.
\label{407}
\end{equation}
This value of $\rho$
is different from the value of $\xi$
in Eq.~(\ref{119}),
albeit of the same order of magnitude.
Therefore,
the energies and momenta of the massive neutrino contributions
to the flavor transition amplitude (\ref{175})
are different from the average energies and momenta of the massive neutrino
components of the neutrino state created in the production process.

From Eq.~(\ref{174}) we obtain
\begin{equation}
\omega
\simeq
\frac
{
\lambda_D
+
\left(v_D^x-1\right)^2+(v_D^y)^2+(v_D^z)^2
}
{
\lambda_D
+
(v_D^y)^2+(v_D^z)^2
}
\simeq
17
\,,
\label{408}
\end{equation}
and from Eq.~(\ref{180})
we obtain the coherence lengths
\begin{equation}
L^{\mathrm{coh}}_{ab}
\simeq
\frac{2 \times 10^{16} \, \mathrm{eV}^2}{|\Delta{m}^2_{ab}|}
\,
\sigma_{xD}
\,.
\label{409}
\end{equation}
If
$|\Delta{m}^2_{ab}| \sim 1 \, \mathrm{eV}^2$
and
$\sigma_{xD}$ is
given approximately by the inter-atomic distance,
$\sigma_{xD} \sim 10^{-8} \, \mathrm{cm}$,
we have
$
L^{\mathrm{coh}}_{ab}
\sim
10^3 \, \mathrm{km}
$,
a sufficiently long coherence distance
for short-baseline neutrino oscillation experiments.

From Eq.~(\ref{472})
we find that $\zeta$ is approximately given by
\begin{equation}
\zeta
\simeq
\frac
{\lambda_D}
{
\lambda_D
+
\left( 1 - \vet{\ell} \cdot \vet{v}_D \right)^2
}
\simeq
0.53
\,,
\label{410}
\end{equation}
and the condition (\ref{1752}) is satisfied for
\begin{equation}
m_a^2 \ll \left( 10^{7} \, \mathrm{eV} \right) \sigma_{pD}
\,.
\label{511}
\end{equation}
If
$\sigma_{xD}$ is
given approximately by the inter-atomic distance,
we have
$\sigma_{pD} \sim 10^{3} \, \mathrm{eV}$
and the inequality (\ref{411}) reads
$m_a^2 \ll 10^{10} \, \mathrm{eV}^2$,
that is certainly satisfied.

\section{Conclusions}
\label{Conclusions}

We have presented a quantum-field-theoretical model of
neutrino oscillations in which
the neutrino propagating between the production and detection
processes is described by the wave-packet state (\ref{099}),
which is determined by the production process
as naturally expected from causality.
Since in real oscillation experiments neutrinos
propagate over macroscopically large distances,
we think that this model is preferable
over the quantum-field-theoretical model
of neutrino oscillations with virtual intermediate neutrinos
presented in Refs.~\cite{Giunti-Kim-Lee-Lee-93,Giunti-Kim-Lee-Whendo-98}.

We have considered production and detection processes
of the form (\ref{prod}) and (\ref{dete})
(other types of production and detection processes
can be considered with straightforward
changes to the formalism)
in which the interacting particles are described by localized wave packets.
The Gaussian form of the wave packets of interacting particles
that we have adopted
should be considered as an approximation of the real
wave packets.
In any case,
other wave packets which are sharply peaked around their average momentum
lead to the same results,
because these results depend on several integrations
that are calculated with the saddle-point approximation.

We have obtained the flavor transition probability,
presented in Eq.~(\ref{178}),
that is almost identical to the one derived
in the quantum-field-theoretical model
with virtual intermediate neutrinos
\cite{Giunti-Kim-Lee-Whendo-98}.
Only the quantity $\omega$, given in Eq.~(\ref{174}),
is somewhat different from the corresponding one
in Ref.~\cite{Giunti-Kim-Lee-Whendo-98}.

As already discussed in Refs.~\cite{Giunti-Kim-Lee-Whendo-98,Beuthe:2002ej},
the flavor transition probability
is determined not only by the production process,
but also by the detection process
\cite{Kiers-Nussinov-Weiss-PRD53-96}.
In particular,
we have shown that the energies and momenta
of the massive neutrino components
relevant for the oscillations
are in general different from the average energies and momenta
of the massive neutrino components
of the propagating neutrino state,
which are determined only by the production process.

Our result confirms the correctness of
the standard expression (\ref{179}) for the oscillation lengths
of extremely relativistic neutrinos
and the existence of coherence lengths
given by Eq.~(\ref{180}).
We agree with the author of Ref.~\cite{Beuthe:2002ej}
on the possibility to extend without limit
the coherence length with an appropriate setup of the detection
process.
We have shown that
the coherence length cannot be increased
without limit by decreasing the momentum
uncertainty of the detection
(or production)
process,
because a vanishing momentum uncertainty
does not allow the coherent detection
(or production)
of different massive neutrino components.
Instead,
the coherence length can be increased
without limit
by choosing a detection
(or production)
process such that $\rho$, given in Eq.~(\ref{161}), is zero and
$\omega$, given in Eq.~(\ref{174}), is infinite.
As discussed for the example presented in
Subsection~\ref{Dete: Equal energy limit},
in practice
the possibility to
increase without limit the coherence length
is rather unrealistic.
As shown in Subsection~\ref{Dete: Realistic case},
in a realistic experimental setup
the coherence length is very long,
but not infinite.

Finally,
let us recall the important remark
presented at the end of the introductory Section~\ref{Introduction}.
In this paper we have assumed that the particles
$ P_I $, $ P_F $ and $ \ell^{+}_{\alpha} $
participating to the neutrino production process (\ref{prod})
are described by the pure wave-packet states (\ref{states})
in which all their properties are determined.
In practice it is common that the knowledge
of these properties is less than complete.
In this case the particles
$ P_I $, $ P_F $ and $ \ell^{+}_{\alpha} $
must be described by statistical operators (density matrices)
constructed from appropriate incoherent mixtures of the
pure wave-packet states (\ref{states}).
Consequently,
the neutrino created in the production process
must be described by a statistical operator
constructed from an incoherent mixture
of the pure wave-packet states (\ref{099}).
Similarly,
if there is incomplete knowledge of the properties of the particles
$ P_I $, $ P_F $ and $ \ell^{+}_{\alpha} $
and the particles
$ D_I $, $ D_F $ and $ \ell_{\beta}^{-} $
participating,
respectively,
to the production and detection processes,
the oscillation probability is given by an appropriate
average of the probability in Eq.~(\ref{178})
over the unknown quantities.

\begin{acknowledgments}
I would like to thank S.M. Bilenky for stimulating discussions
and enlightening remarks,
and C.W. Kim for a long and fruitful collaboration
on the study of the theory of neutrino oscillations.
I am specially indebted with M. Beuthe
for his remarks concerning the energy uncertainty
of wave packets,
that helped to correct several wrong equations
in the first version
of the paper appeared in the electronic archive hep-ph.
\end{acknowledgments}

\end{document}